\documentclass[pre,twocolumn,aps,reprint,amsmath,amssymb,longbibliography]{revtex4-1}% APS journal style
\usepackage{graphicx}% Include figure files
\usepackage{amsmath}
\usepackage{dcolumn}% Align table columns on decimal point
\usepackage{bm}% bold math
\usepackage{hyperref}% add hypertext capabilities
\usepackage[]{amssymb}
\usepackage{color}
\usepackage{epstopdf}
\usepackage{mathscinet}
\newcommand{\ER}{Erd\H{o}s--R\' enyi{}}

\begin{document}

\title{Hipsters on Networks: How a Minority Group of Individuals Can Lead to an Anti-Establishment Majority}
\date{\today}
\author{Jonas S. Juul}
	\email{jonas.juul@nbi.ku.dk}
\affiliation{Niels Bohr Institute, University of Copenhagen, Blegdamsvej 17, Copenhagen 2100-DK, Denmark}
\author{Mason A. Porter}
	\email{mason@math.ucla.edu}

\affiliation{Department of Mathematics, University of California, Los Angeles, Los Angeles, California 90095, USA}  
\affiliation{Oxford Centre for Industrial and Applied Mathematics, Mathematical Institute, University of Oxford, Oxford OX2 6GG, UK}
	\affiliation{CABDyN Complexity Centre, University of Oxford, Oxford OX1 1HP, UK}

%%%%%%%%

\begin{abstract}

The spread of opinions, memes, diseases, and ``alternative facts'' in a population depends both on the details of the spreading process and on the structure of the social and communication networks on which they spread. One feature that can change spreading dynamics substantially is heterogeneous behavior among different types of individuals in a social network. In this paper, we explore how \textit{anti-establishment} nodes (e.g., \textit{hipsters}) influence the spreading dynamics of two competing products. We consider a model in which spreading follows a deterministic rule for updating node states (which describe which product has been adopted) in which an adjustable fraction $p_{\rm Hip}$ of the nodes in a network are hipsters, who choose to adopt the product that they believe is the less popular of the two. The remaining nodes are conformists, who choose which product to adopt by considering which products their immediate neighbors have adopted. We simulate our model on both synthetic and real networks, and we show that the hipsters have a major effect on the final fraction of people who adopt each product: even when only one of the two products exists at the beginning of the simulations, a very small fraction of hipsters in a network can still cause the other product to eventually become the more popular one. To account for this behavior, we construct an approximation for the steady-state adoption fraction on $k$-regular trees in the limit of few hipsters. Additionally, our simulations demonstrate that a time delay $\tau$ in the knowledge of the product distribution in a population, as compared to immediate knowledge of product adoption among nearest neighbors, can have a large effect on the final distribution of product adoptions. Using a local-tree approximation, we derive an analytical estimate of the spreading of products and obtain good agreement if a sufficiently small fraction of the population consists of hipsters. In all networks, we find that either of the two products can become the more popular one at steady state, depending on the fraction of hipsters in the network and on the amount of delay in the knowledge of the product distribution. Our simple model and analysis may help shed light on the road to success for anti-establishment choices in elections, as such success --- and qualitative differences in final outcomes between competing products, political candidates, and so on --- can arise rather generically in our model from a small number of anti-establishment individuals and ordinary processes of social influence on normal individuals.
 
\end{abstract}

%%%%%%%%%%


%Use showkeys class option if keyword
\maketitle

%%%%%%%

%%%%%%

\section{Introduction}

The study of spreading phenomena on networks has received considerable attention in many disciplines, including sociology, economics, physics, biology, computer science, mathematics, and others \cite{lehman-ahn-book,porter2016,fowlerreview,yamir-jcn2013,granovetter78,valente-book,jackson2013,kkt2003,watts2002,loreto2009,Christakis07,centola2007,dodds2005,aral2009,ugander2012,goel-preprint,monsted2017evidence,mollison1977}. In analogy with the spread of infectious diseases in populations of susceptible individuals, the spread of social phenomena (such as opinions, actions, memes, information, misinformation, and alternative facts) is often viewed as a contagion process that spreads through a network's nodes, which are connected to each other via one or more types of edges. The nodes can represent entities such as people or institutions \cite{10.1257/aer.99.5.1899,Centola2005_norms,elsinger_risk} (or other things); and the edges can represent physical proximity, communication channels, sociological interactions (e.g., different types of relationships), or something else. An important goal of many studies of the spread of social contagions is the identification of criteria that determine when the phenomenon that is spreading reaches a large fraction of a population or subpopulation \cite{porter2016,lehman-ahn-book}.

Scholars have used various approaches for studying contagions on networks. These include game theory \cite{NWS:8888780}, statistical physics\cite{loreto2009}, agent-based models \cite{juul2011locally}, and systems of coupled differential equations or stochastic processes \cite{porter2016, Dodds04, miller2017, MelnikChaos13, perez2011synergy, juul2017synergistic}. The temporal dynamics and peak size of an outbreak are influenced both by the specific model of a contagion and by the structure of the network on which it spreads \cite{rom-review2015, taylor2015, colizza-prx2015, gleeson2013PRX, gleeson2008, centola2007, gleeson-watts-weighted, MelnikChaos13, PhysRevE.88.012818, Centola2005_norms,juul2017synergistic}. One of the focal ideas is to examine when a disease or idea, initially carried by a small fraction of nodes, can become widespread in a network. When a large fraction of a population or subpopulation becomes infected (or adopts an idea), one says that a \textit{cascade} has occurred, and cascading phenomena have been studied in a wide variety of systems, ranging from financial networks \cite{elsinger_risk} to social media like Twitter \cite{PhysRevX.6.021019}. For example, a failing financial institution can cause a cascade of failures of numerous other financial institutions, a tweet can result in a cascade of tweets that promotes the opinion of the original tweeter (perhaps influenced by the actions of `bot' or sockpuppet accounts \cite{monsted2017evidence}), and widely-spread alternative facts can influence the opinions of a large population of voters \cite{fakenews2017}. Notwithstanding these dystopian examples, cascading behavior can beneficial, neutral, or harmful.

Models of cascading behavior on networks can have either stochastic or deterministic state-update rules, and the update rules in most models only consider nodes that are adjacent to the node under consideration. One can of models that traditionally have deterministic update rules are \textit{threshold models} for social contagions. The simplest example is the Watts threshold model (WTM) \cite{watts2002,valente-book,granovetter78}, a type of bootstrap percolation \cite{chalupa1979}, in which each node is assigned a threshold from some distribution. When considering a node for updating, if the fraction of its neighbors that are adopters is at least as large as its threshold, it becomes an adopter itself. There are numerous variants and generalizations of the WTM, including ones with adoption thresholds that are based on the number (rather than the fraction) of adopted neighbors \cite{centola2007,cent-eguil}, ones with multiple adoption stages \cite{MelnikChaos13}, ones with ``synergy'' from other nearby adopters \cite{juul2017synergistic}, and ones with timers in addition to adoption thresholds \cite{oh2018}.

Efforts to develop mathematical models for the spread of products or innovations date at least as far back as the 1960s. Rogers \cite{rogers2010diffusion} described qualitatively (as a sigmoidal shape) how the number of adopters should look as a function of time. Bass \cite{bass1969new} developed a model for the adoption of innovations that was inspired by models for biological contagions. Bass's model results in sigmoidal-shaped adoption curves, and it has been generalized in various ways \cite{bass1980relationship,nevers1972extensions,bass1994bass,bass2004comments}. More recent studies have considered models in which agents of different types can have significant effects on the final distributions of products or innovations in a population. For example, \cite{gordon2016adoption} showed that temporal cycles of adoption can occur if some nodes are allowed to regret adopting an innovation while other (``contrarian'') nodes resist adopting innovations. References \cite{dodds2013limited,harris2014} found rich behavior (including chaotic dynamics) in a social contagion model that incorporates an aversion to complete conformity. 

Contrarian agents, a key aspect of the present paper, have been incorporated into various types of models of opinion dynamics and hierarchy formation. In the 1980s, \citet{galam1986} illustrated a hierarchical mechanism that allows a minority community to elect its preferred candidate instead of that of the majority. Galam and collaborators have also examined the effects of contrarian \cite{galam2004} and stubborn \cite{galam2007,galam2016} agents on opinion dynamics (though typically without any network structure). \citet{nyczka2012,nyczka2013} studied various opinion models (e.g., $q$-voter models) on a complete graph to highlight an important distinction between two types of nonconformity --- anti-conformity and independence --- that have distinct implications for social dynamics. \citet{khalil2018noisy} incorporated contrarians into a noisy voter model, and they illustrated that a few contrarians can substantially alter the dynamics of the model. \citet{apriasz2016} examined an opinion model that includes ``snobs'', who conform to nodes in their own community but anti-conform to nodes in others, to examine how the density of connections between two communities can effect phenomena such as fashion cycles. One can also consider contrarian individuals in the context of economic markets, such as in work by Sznajd-Weron and Weron \cite{szajd2003}, who studied an Ising model on a rectangular lattice to model advertising in duopoly markets. More recent work related to contrarian agents, in addition to \cite{dodds2013limited,harris2014,gordon2016adoption}, includes that of \citet{mellor2015influence}, who examined a population in which nodes can either adopt a product or become ``luddites'', who oppose the spread of innovation. They found that luddites greatly limit adoption if the adoption rate is high but not if it is low. \citet{gambaro2017} illustrated that contrarian agents can be a source of disorder in opinion dynamics, and Ferrara and collaborators have investigated how individual social-media accounts controlled by bots can exert a considerable influence on political elections and social cascades \cite{ferrara2017disinformation,monsted2017evidence,bessi2016social}.

Anti-conformity can manifest in a variety of ways in society. For example, it has been reported in the sociological literature that partisan bias can result in some groups of individuals misinterpreting data and explanations of experts (e.g., with respect to the issue of climate change), in conflict with the intended message, such that it fits with the personal beliefs of the group \cite{guilbeault2018social,jamieson2014leveraging}. In a recent example about information spreading, \citet{petersen2018need} provided psychological assessment of motivations to share hostile political rumors (e.g., in the form of ``fake news'') among citizens of democratic societies, concluding that such rumors are often shared by individuals because they believe it can mobilize their audience against a disliked establishment (rather than because they think that these rumors are true).

In the present paper, we examine the influence of \emph{anti-establishment} nodes, such as \emph{hipsters}, on spreading processes in a social network. Individuals who specifically prefer something other than the established standard in society have manifested in several ways over the last decade. They include members of anti-establishment movements in Western Europe and the United States of America, who have hugely impacted the geopolitical landscape, to the curious style of hipsters in cities throughout the world. 
In some cases, such as the 2016 ``Brexit'' vote \cite{hobolt2016brexit} and the 2016 American presidential election \cite{oliver2016rise}, anti-establishment opinions appear to have spread to so many people that they exerted a major influence on political outcomes. In this paper, we ask the following questions: (1) How does a large fraction of a population decide to choose something different from the established standard? (2) How can a small fraction of individuals spread their anti-establishment opinions to a majority (or at least to a very large minority) of the rest of a population? (3) Can we capture these ideas using a simple mathematical model of a spreading process on a network?

A few years ago, a statistical-physics approach was used to examine how anti-conformists (i.e., hipsters) who make decisions against the majority, thereby attempting to stand out from the crowd, may all end up ``looking the same'' \cite{touboul2014hipster} (wearing the same clothes, buying the same products, having the same opinions, and so on). This study observed that the dynamics of a population was influenced greatly by delays in the knowledge of hipsters and by how large a fraction of the population are hipsters.  In the model in \cite{touboul2014hipster}, individuals interact with their environments and switch between two states $\{-s_i,s_i\}$ with a probability that depends on this environment and on whether an individual is conformist (preferring to be aligned with the environment) or a hipster (preferring to be opposite to its environment). The model has a phase transition, which determines whether or not hipsters ultimately attain the same state. Touboul \cite{touboul2014hipster} referred to the anti-conformists with delayed knowledge as ``hipsters''; because the model that we introduce in this study includes anti-conformists with delayed knowledge of the global product distribution, we also adopt this terminology. Word choice notwithstanding, our approach, focus, and type of model --- which builds on threshold models for social contagions --- are rather different from those in \cite{touboul2014hipster}. 

We will explore how anti-conformists (i.e., hipsters) affect the spreading of competing products in a network by generalizing the WTM to a network with two types of nodes --- hipsters and conformists --- who respond differently to adoptions. Conformists prefer to adopt products (or memes, opinions, messages, etc.) that the majority of their neighbors have adopted at time $t-1$. Hipsters, however, prefer to adopt the product that is the less popular of the two products at some previous time $t- \tau$. Their choice to adopt a product uses the same adoption condition as with the conformists. This is a strong assumption, and we make it partly for simplicity (as it allows us to build from the WTM) and partly because it reflects a scenario in which an anti-conformist may more actively rebel on an issue that is sufficiently established in its neighborhood in a network. In our model, both conformists and hipsters first choose to buy some product or form an opinion, and then they choose which one to adopt. In their study of the effect of luddites, \citet{mellor2015influence} assumed that the probability of a node becoming a luddite is proportional to the rate of change in the density of adopters of its neighbors. This resembles our choice that a node's neighborhood influences whether or not it chooses to adopt a product, bit it differs from the fact that our nodes are either inherently a conformist or inherently a hipster. The delay $\tau$ in our model encodes the fact that knowledge about the total population is not instantly available; instead, it is collected over some time before it is available. See Gleeson et al. \cite{gleeson2014} for a model, without network structure, that illustrates another type of competition between local information (in the form of a social-media feed) and global information (in the form of a best-seller list). 

The rest of our paper is organized as follows. In Section \ref{sec:Model}, we introduce our model for spreading on networks under the influence of hipster nodes. In Section \ref{sec:FB_realization}, we examine our model on a Facebook network. In Section \ref{sec:Analysis}, we develop an analytical approximation to describe the time-dependent fraction of nodes who adopt the products as a function of their degree $k$ and an adoption threshold $\phi$. In Section \ref{sec:Realization}, we examine our model on several classes of empirical networks and investigate the final fractions of nodes who adopt the two products as a function of the time delay $\tau$ and hipster fraction $p_{\mathrm{Hip}}$. In Section \ref{sec:regular_trees}, we explain the observed behavior in the limit of few hipsters, and we obtain an 
approximation for the fraction of nodes who adopt the different products as a function of the number of hipsters. We conclude and discuss our results in Section \ref{sec:Conclusions}.

%%%%

\section{A threshold model with hipsters}\label{sec:Model}

A popular type of model for spreading processes on networks are \textit{threshold models} of social influence\cite{valente-book,porter2016,lehman-ahn-book}. To set up a simple example of a threshold model, consider a network with $N$ nodes, and suppose that each node $i$ has an independently-assigned threshold $\phi_i$ that we draw from a distribution $f(\phi)$. We also suppose that a node can be in one of two states: \textit{active} or \textit{inactive}. An active node has adopted a product (or meme, opinion, etc.) that is spreading through a population, and an inactive node has not adopted the product. (We will use the term ``product'' from now on.) Once a node becomes active, it stays active forever. The threshold of a node determines how difficult it is to activate that node, so one can construe a node's threshold value as its stubbornness level. Node $i$ becomes active if a peer pressure, which in the WTM is equal to the fraction of active nodes among $i$'s neighbors, is greater than or equal to its threshold $\phi_i$.

We seek to develop a model for competing products that spread in a population that includes hipsters. Therefore, in our model, each node $i$ has a value $H_i\in \{0,1\}$, such that $H_i=0$ indicates that node $i$ is a \textit{conformist} and $H_i=1$ indicates that node $i$ is a \textit{hipster}. We update nodes synchronously. At each discrete time $t \geq 1$, we assume that conformists know the distribution of products among their immediate neighbors at the previous time step $t'=t-1$, whereas hipsters know the distribution of products in the total population at an earlier time step $t_\tau = t-\tau$ (where $\tau \in \mathbb{N})$. The first updating step occurs at $t=1$. If $t-\tau < 0$, we let $t_\tau = 0$.

\begin{figure*}[tb]
\centering
\includegraphics[width= \linewidth]{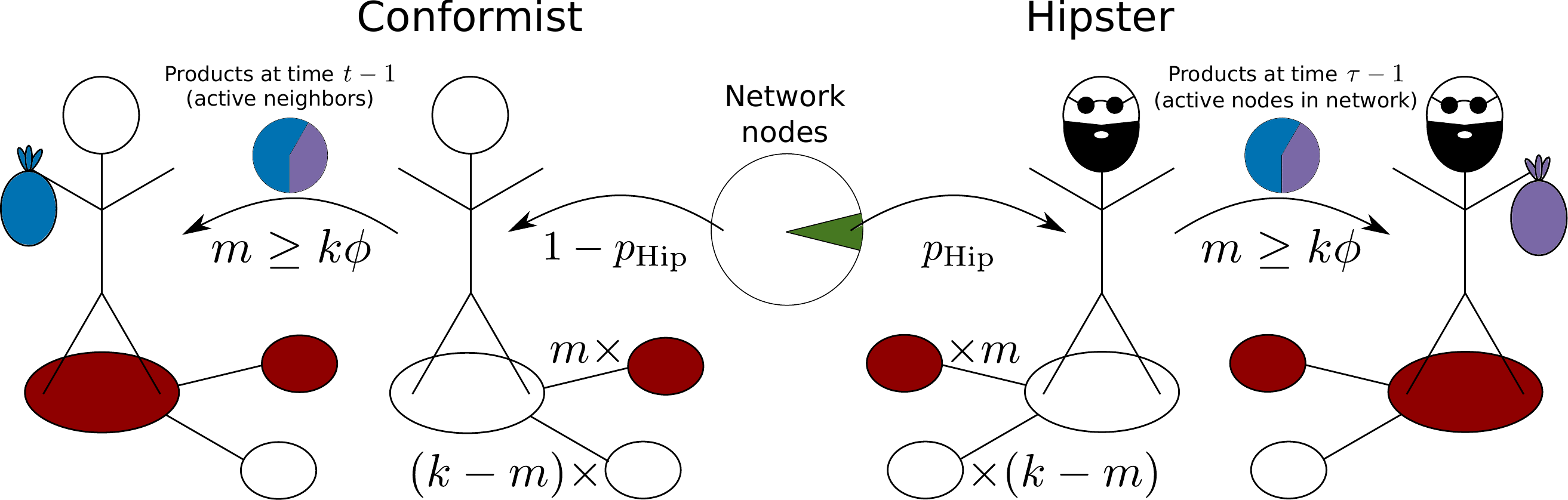}
\caption{Illustration of our model of a threshold-based social contagion with hipsters. A node is a hipster with probability $p_{\mathrm{Hip}}$, and it is a conformist with probability $1-p_{\mathrm{Hip}}$. If at least a fraction $\phi_i$ of the neighbors of node $i$ are active (as indicated by the red coloring) at discrete time step $t-1$, the node activates and adopts a product at time step $t$ (for $t \geq 1$). We then need to consider which products have been adopted by node $i$'s neighbors and the relative popularity of different products in the whole network. If node $i$ is a conformist, it adopts the product that the majority of its active neighbors have adopted at time $t-1$. However, if node $i$ is a hipster, it adopts the product that is less popular among the active nodes in the network at time step $t-\tau$ (where $\tau\in\mathbb{N}$). For both node types, a tie results in a node choosing one of the two products (blue versus purple) with equal probability.}
\label{fig:The_model}
\end{figure*}

A node chooses to adopt a specific product in two steps. First, the node must become active, as determined by whether sufficiently many of its neighbors are active. If the fraction of neighbors that are active at time $t-1$ is at least as large as the node's threshold, it becomes active at time $t$. If node $i$ becomes active, it immediately adopts one of two possible products. If $H_i=0$, node $i$ is a conformist and thus adopts the product that most of its active neighbors have adopted at time step $t-1$. However, if $H_i=1$, node $i$ is a hipster and thus adopts the product that is the less popular in the total population at time $t_\tau = t-\tau$. For both $H_i$ values, a tie results in the node choosing one of the two products with equal probability. Each node can adopt only a single product; and once it has adopted a product, it never switches to the other product or becomes inactive. To keep track of the product distribution, we associate a variable $S_i$ to each node $i$. If $S_i=0$, node $i$ is inactive; if $S_i = 1$, node $i$ has adopted product $A$; and if $S_i = 2$, node $i$ has adopted product $B$. We summarize our model and the decision process in Fig.~\ref{fig:The_model} and its caption. We have made a {\sc Python} script to simulate our model available at \url{https://sid.erda.dk/share_redirect/hZOCeo4qU5}.
  
At $t=0$, we activate a single node with product $A$, and we introduce product $B$ when the first hipster chooses to adopt a product.  In principle, it is possible to generalize our model to consider arbitrarily many products spreading on a network, but we consider only the case of two products for simplicity.

Although it may seem somewhat artificial that the conformists in our model use only local information to decide which product to adopt, whereas the hipsters use only global information, it is both convenient and illustrative to generalize the WTM model by adding one specific feature because it so well-studied. This is also appropriate for exploring the competition between local and global forms of influence. We examine our hipster model on both synthetic networks and empirical social networks. Our main goal is to examining whether (and when) a small fraction $p_{\mathrm{Hip}}$ of hipster nodes can lead to a majority of a network's nodes adopting a less-popular product at the beginning of a spreading process. We find that the fraction of nodes that adopt the insignificant product depends in an interesting way on the delay $\tau$ in the hipsters' knowledge of the product distribution in the total population.

%%%%%

\section{Simulation of our model on a Facebook network}\label{sec:FB_realization}

We start by simulating our model on the {\sc Northwestern25} network from the {\sc Facebook100} data set \cite{traud_social_2012}. This network consists of the friendship relationships on Facebook at Northwestern University on one day in autumn 2005. The network has 10537 nodes, a mean degree of $\langle k \rangle \approx 92$, and a maximum degree of $k_{\mathrm{max}} = 2105$. We assign a threshold of $\phi = 1/33$ to each node. In addition to this threshold, we independently assign each node a value $H_i \in \{0,1\}$ with some hipster probability $p_{\mathrm{Hip}}$ to be $H_i = 1$. Therefore, different simulations of our model with a specified hipster probability do not in general have the same number of hipster nodes. We examine our model with two different time delays and two different values for the fraction of hipsters in the network. We consider $\tau = \{ 1,4\}$ and $p_{\mathrm{Hip}} = \{0.04, 0.30 \}$, and we conduct simulations for the four combinations of these parameter choices. 

\begin{figure}
\includegraphics[width=\linewidth]{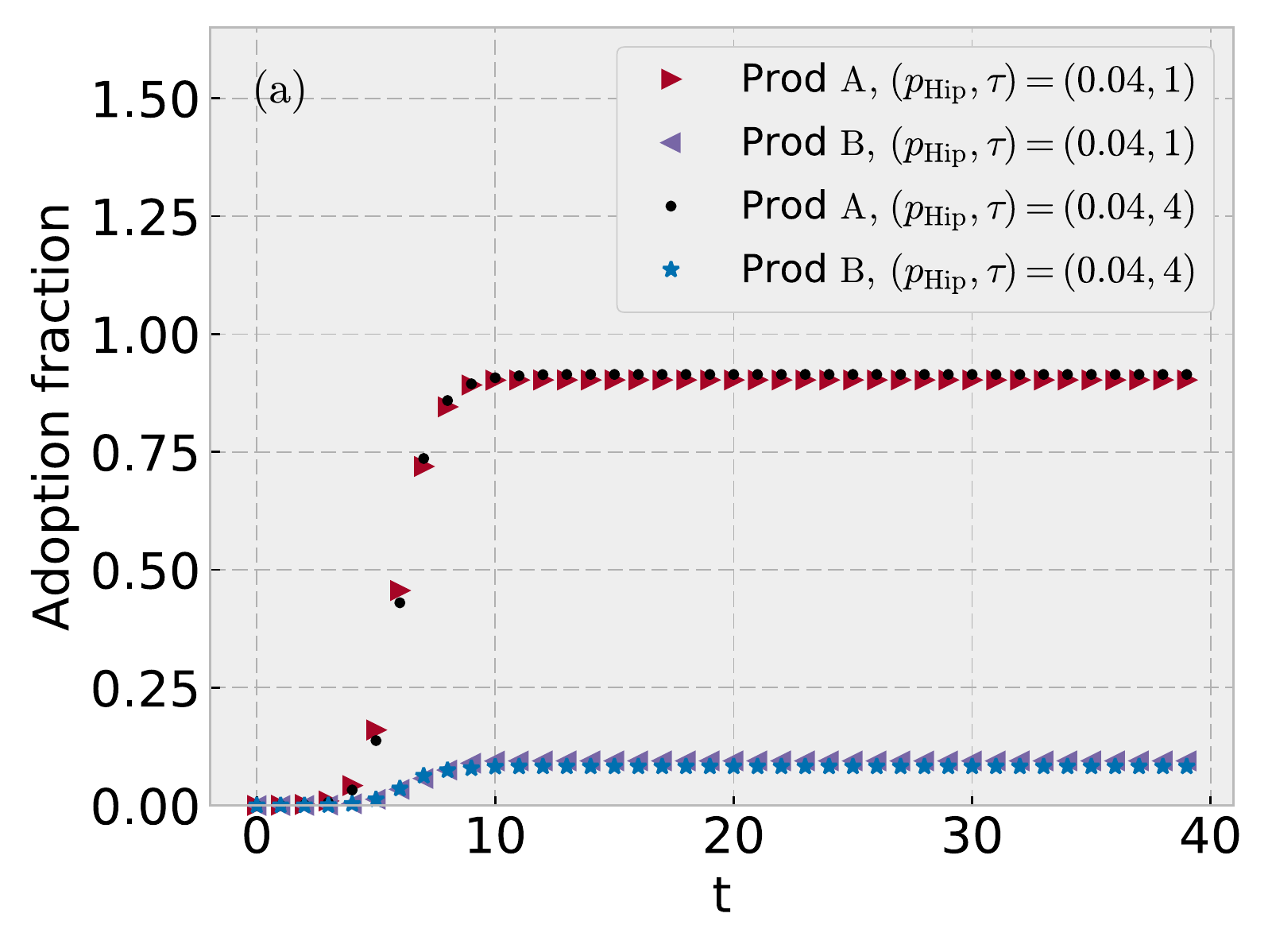}
\\
\includegraphics[width=\linewidth]{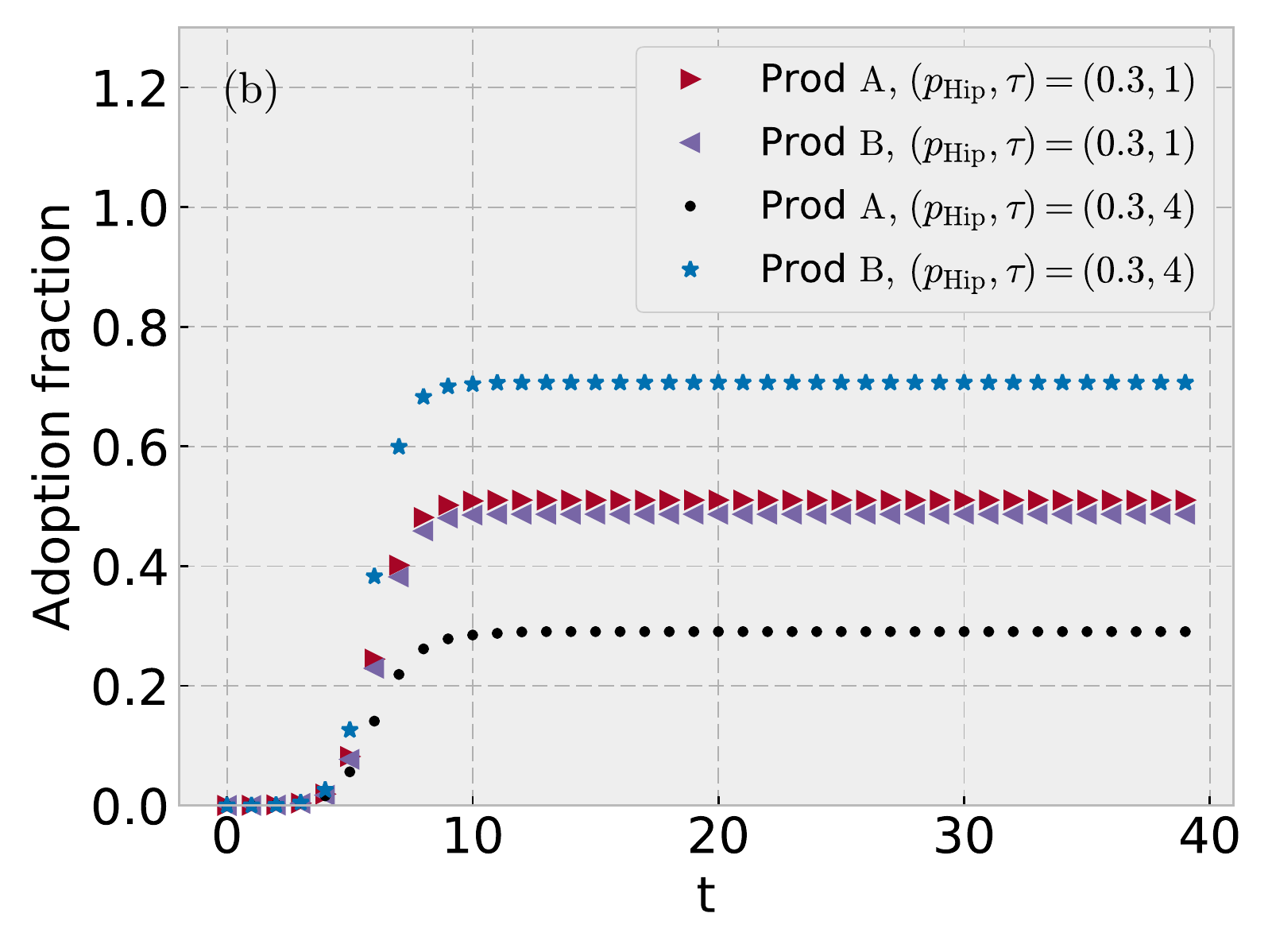}
\caption{Example of the behavior of our hipster threshold model on a Facebook network for delay values $\tau = \{ 1,4\}$ for (a) a hipster fraction of $p_{\mathrm{Hip}}=0.04$ and (b) a hipster fraction of $p_{\mathrm{Hip}}=0.30$. In each panel, we show the fraction of nodes in the network that have adopted each of the two products as a function of time. For a given value of the delay, the curves are almost indistinguishable from each other. In panel (b), the adoption fractions of the two products are clearly different when we use different delays ($\tau = 1$ and $\tau = 4$). For $\tau = 1$, the final fractions that have adopted product $A$ and $B$ are approximately equal. However, for $\tau = 4$, product $B$ is ultimately adopted more than product $A$. For both panels, each data point is a mean over $200$ simulations on the same network (the {\sc Northwestern25} network of the {\sc Facebook100} data set\cite{traud_social_2012}), where we choose both seed nodes and hipster nodes uniformly at random for each of the simulations.
}
\label{fig:FB_single_parameters}
\end{figure}

For each parameter pair, we choose a single node uniformly at random and suppose that it has adopted product $A$ at time $t = 0$. This node acts as a seed for the spreading process on the network. We introduce another product, labeled $B$, when the first hipster node is activated. Thus, product $B$ will never be adopted by any node if the network has no hipsters; and product $A$ has a head start when product $B$ is adopted for the first time. We stop our simulations after the dynamics reaches a steady state, in which no further adoptions occur. At each time step, we track the fraction of the nodes that have adopted each of the two products. We conduct $200$ simulations --- each with a seed chosen uniformly at random, and determining new hipster nodes for each simulation --- and we average the fraction of nodes that have adopted each product at time step $t$ over these $200$ simulations. We show the results of these simulations in Fig.~\ref{fig:FB_single_parameters}.

In Fig.~\ref{fig:FB_single_parameters}(a), we plot the fraction of nodes in the adopted state for each of the products at time $t$ for simulations in which the hipster fraction is $0.04$. For these parameters, the curves are indistinguishable for the two values of the delay time $\tau$. A much larger fraction of nodes adopts product $A$ than product $B$. In Fig.~\ref{fig:FB_single_parameters}(b), we show the corresponding curves for simulations in which the hipster fraction is $0.30$. The results for different delay times $\tau$ are now clearly distinguishable. For $\tau = 1$, the fraction of nodes that have adopted the two products are approximately equal; for $\tau = 4$, however, the fraction of nodes that have adopted product $B$ is much larger than the fraction that have adopted product $A$.

%%%%%

\section{Analysis}\label{sec:Analysis}

We approximate the temporal spreading of products on a network using a pair approximation (as in \cite{gleeson2008,MelnikChaos13,juul2017synergistic,gleeson07}) that relies on the hypothesis that the network is locally tree-like \cite{localtreeapprox_mason}.\footnote{Pair approximations have also been employed in other types of opinion models, such as for a $q$-voter model in \cite{arkad2017}.} Let $\rho_\lambda^{(\phi,k)}(t)$ denote the density of nodes with threshold $\phi$ and degree $k$ that are in the adopted state, for a product $\lambda\in\{A,B\}$, at time step $t$. We write the recursion relation
%\begin{onecolumn}
\begin{widetext}
\begin{align}
	\begin{split}
\rho_\lambda^{(\phi,k)}(n+1) &= \rho_\lambda^{(\phi,k)}(n)+(1-\rho_\lambda^{(\phi,k)}(n))\sum_{k'=1}^{k}F(k,k',\phi)\\
	& \times B_{k'}^{k}\left(\sum_{\beta=1}^m\bar{q}_\beta^{(\phi,k)}(n)\right) 
 \left[(1-p_{\mathrm{Hip}})\sum_{k''>\frac{k'}{2}}^{k'} B_{k''}^{k'}\left(\frac{\bar{q}_\lambda^{(\phi,k)}(n)}{\sum_{\beta=1}^m\bar{q}_\beta^{(\phi,k)}(n)}  \right) \right. \\
	& +  \left. p_{\mathrm{Hip}} \prod_{\beta\ne \lambda} \Theta\left(\sum_{k,\phi}\rho^{(\phi, k )}_\lambda(n+1-\tau) - \sum_{k,\phi}\rho^{(\phi, k )}_\beta(n+1-\tau)\right) \right] \,,
	\end{split}
\label{eq:analytical_approx}
\end{align}
\end{widetext}
%\end{onecolumn}
where $\bar{q}_\beta^{(\phi, k)}(n)$ is the probability that a neighbor, chosen uniformly at random, of an inactive node with threshold $\phi$ and degree $k$ is active and has adopted product $\beta\in\{1,2\}$; the ``response function'' $F(k,k',\phi) = 1$ if $k'/k \ge \phi$ and $F(k,k',\phi) = 0$ otherwise; $\Theta(x)$ is the step function (so it equals $1$ for $x>0$ and $0$ otherwise); and
\begin{equation}
	B^k_l(p) = \binom{k}{l}p^l(1-p)^{k-l}
\end{equation}
is the binomial function for probability $p$. We write $\bar{q}_k^{(\phi,k)}(n)$ as a function of $q_{i}^{(\phi',k')}(n)$, the probability that, for a given inactive node, a neighbor with degree $k'$ and threshold $\phi'$ is active at time step $n$. This probability is given by
\begin{equation}
	\bar{q}_\lambda^{(\phi,k)}(n) = \frac{\sum_{k',\phi'}P\left((k,\phi),(k',\phi')\right)q_{\lambda}^{\phi',k}(n)}{\sum_{k',\phi'}P\left((k,\phi),(k',\phi')\right)}\,,
\end{equation}
where $P\left((k,\phi),(k',\phi')\right)$ is the probability that a node with degree $k$ and threshold $\phi$ is adjacent to a node with degree $k'$ and threshold $\phi'$. Given an active node, the probability that a particular neighbor with degree $k$ and threshold $\phi$ is active is
\begin{widetext}
\begin{equation}\label{eq:qk}
	\begin{split}
	q^{(\phi,k)}_\lambda(n+1) &= \rho_\lambda^{(\phi,k)}(n)+(1-\rho_\lambda^{(\phi,k)}(n))\sum_{k'=1}^{k-1}F(k,k',\phi)\\
	& \times B_{k'}^{k}\left(\sum_{\beta=1}^m\bar{q}_\beta^{(\phi,k)}(n)\right) 
 \left[(1-p_{\mathrm{Hip}})\sum_{k''>\frac{k'}{2}}^{k'} B_{k''}^{k'}\left(\frac{\bar{q}_\lambda^{(\phi,k)}(n)}{\sum_{\beta=1}^m\bar{q}_\beta^{(\phi,k)}(n)}  \right) \right. \\
	& + \left. p_{\mathrm{Hip}} \prod_{\beta\ne \lambda} \Theta\left(\sum_{k,\phi}\rho^{(\phi, k )}_\lambda(n+1-\tau) - \sum_{k,\phi}\rho^{(\phi, k )}_\beta(n+1-\tau)\right) \right] \,.
	\end{split}
\end{equation}
\end{widetext}
The only difference between Eq.~\eqref{eq:qk} and Eq.~\eqref{eq:analytical_approx} stems from the following: in Eq.~\eqref{eq:analytical_approx}, we consider any degree-$k$ node; however, in Eq.~\eqref{eq:qk}, we consider a degree-$k$ neighbor of an inactive node. The latter has a maximum of $k-1$ active neighbors, which is therefore the maximum value of the index of the first sum in Eq.~\eqref{eq:qk}. In these equations, we have assumed that each neighbor of node $i$ is independent of the others, so we are assuming that this process is occurring on a locally tree-like network \cite{localtreeapprox_mason,porter2016}. However, the Facebook network that we used in Section~\ref{sec:FB_realization} has a large local clustering coefficient\cite{traud_social_2012}, so it is not locally tree-like.

%%%%%

\section{Hipster threshold model on synthetic networks}\label{sec:Realization}

We now test our analytical approximations of Section \ref{sec:Analysis} by simulating our model on various synthetic networks with $N = 10,000$ nodes. We assign each node $i$ a threshold $\phi_i$ from some probability distribution $f(\phi)$, which we specify in the following subsections. We also independently assign each node $i$ a value $H_i\in \{0,1 \}$ to determine if it is a hipster. As before, $p_{\mathrm{Hip}}$ denotes the probability of being assigned the hipster value $H_i = 1$. 

As with our simulations on the Facebook network in Section \ref{sec:FB_realization}, we select a single node uniformly at random to have adopted product $A$ at time $t=0$. This node is the seed of the spreading process.
There is a risk that the chosen seed node is located in a neighborhood of very few vulnerable nodes. (A node that can be activated by a single active neighbor is known as a ``vulnerable'' node \cite{watts2002}.)
With such a seed, few nodes are activated in that realization of the dynamical process, and we do not observe a cascade of adoptions in which many nodes adopt a product.

To focus on situations in which many nodes adopt (either of the products), we consider only realizations in which at least some minimal fraction of nodes eventually adopt a product. We take this minimal threshold to be $0.10$. (Another way to examine situations with a lot of spreading is through ``cluster seeding'' \cite{taylor2015}, in which one considers initial conditions in which some node and all of its neighbors start out as adopters.) In Table~\ref{tb:discarded}, we indicate the number of discarded realizations, the mean fraction of adopting nodes in these simulations, and the standard deviation of this number of adopters for several types of networks. The threshold $0.10$ is much larger than the mean fraction of adopters in discarded realizations, and it is much smaller than the fraction of adopting nodes in realizations that we keep. Thus, this choice of threshold entails a clear separation between realizations with cascades of adoption and those without such cascades. In many of our networks, the number of discarded simulations (in which there are few adoptions) is very large, consistent with the empirical study of \citet{goel2012}.

%%%%%%

\subsection{$5$-regular configuration-model networks} \label{five}

We consider configuration-model networks \cite{Fosdick2016} in which every node has degree $5$. As described in \cite{newman2018,Fosdick2016}, we match stubs (i.e., ends of edges) uniformly at random. We suppose that each node has a threshold $\phi_i = 0.19$ with probability $p_0 = 0.8$ and a threshold of $\phi_i = 0.8$ with probability $1-p_0 = 0.2$. Therefore, on average, $80\%$ of the nodes are vulnerable, and $20\%$ of the nodes can adopt only when $4$ or more of their nearest neighbors are adopters. We choose these parameters because there are nodes that need different numbers of neighbors to adopt, and it ensures that a cascade of product adoptions occurs in most realizations of our simulations. 

\begin{figure*} [th!]
\centering
 \includegraphics[width=.43\linewidth]{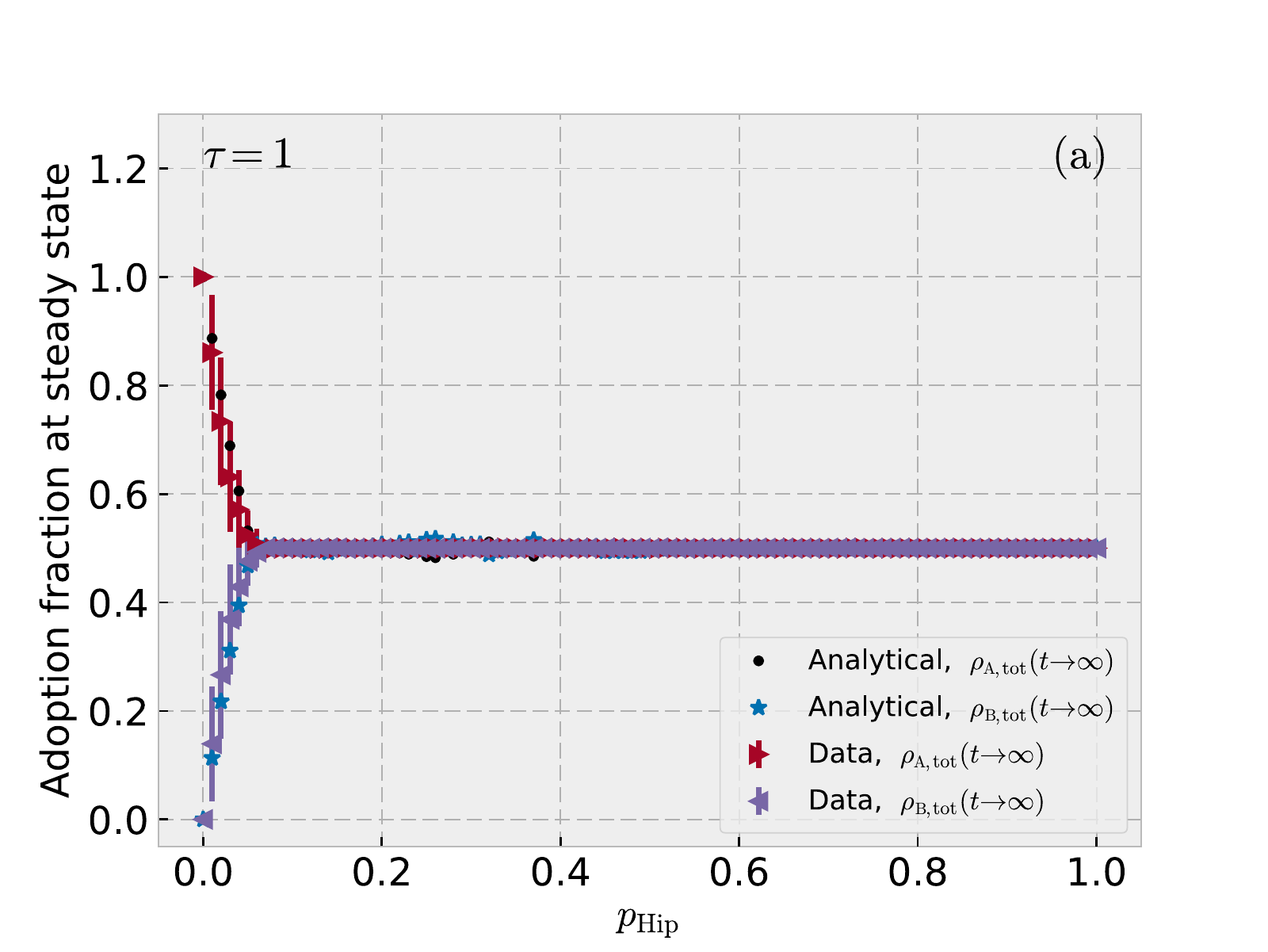}
%\hfill
\includegraphics[width=.43\linewidth]{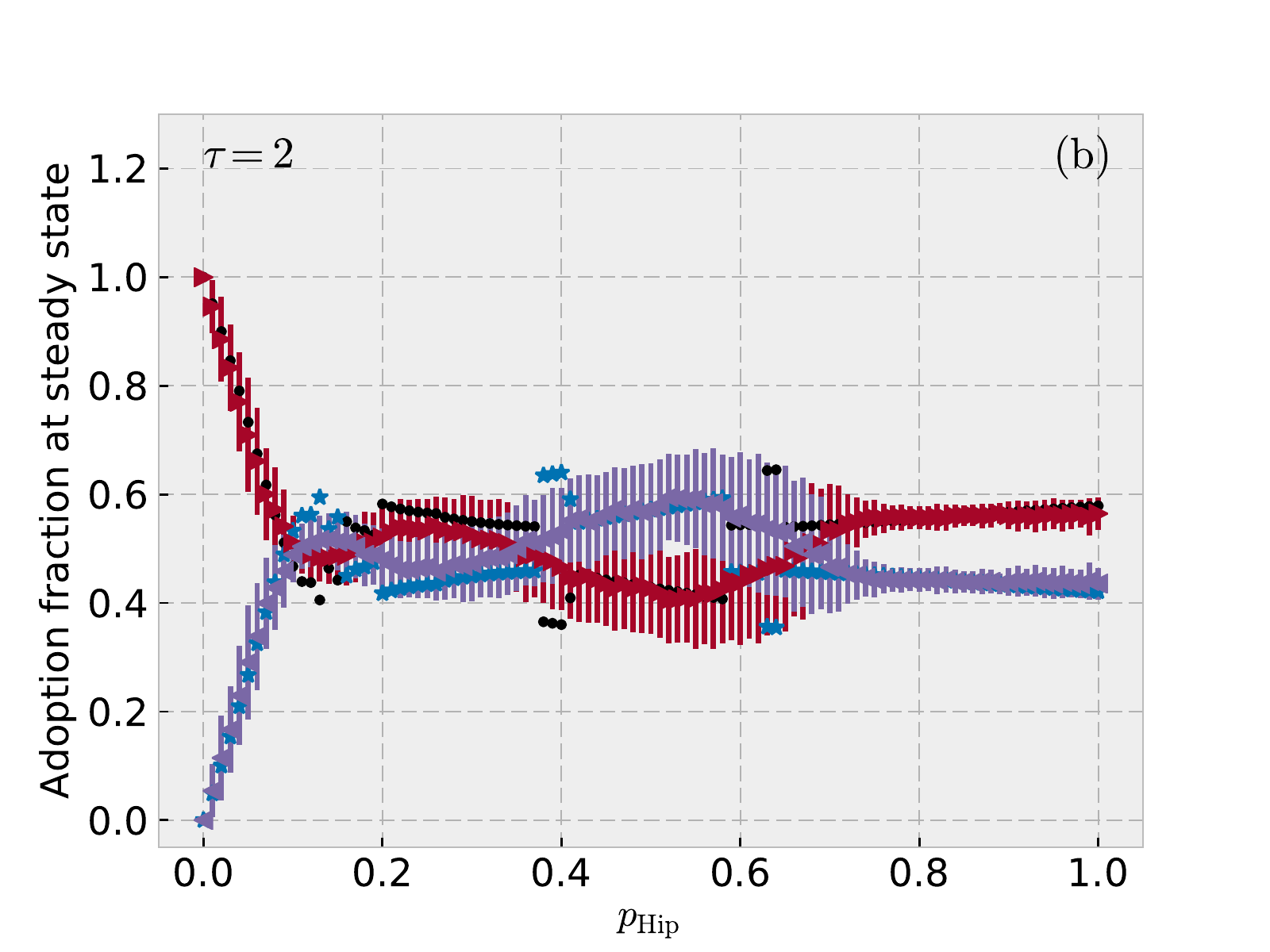}
\\
\includegraphics[width=.43\linewidth]{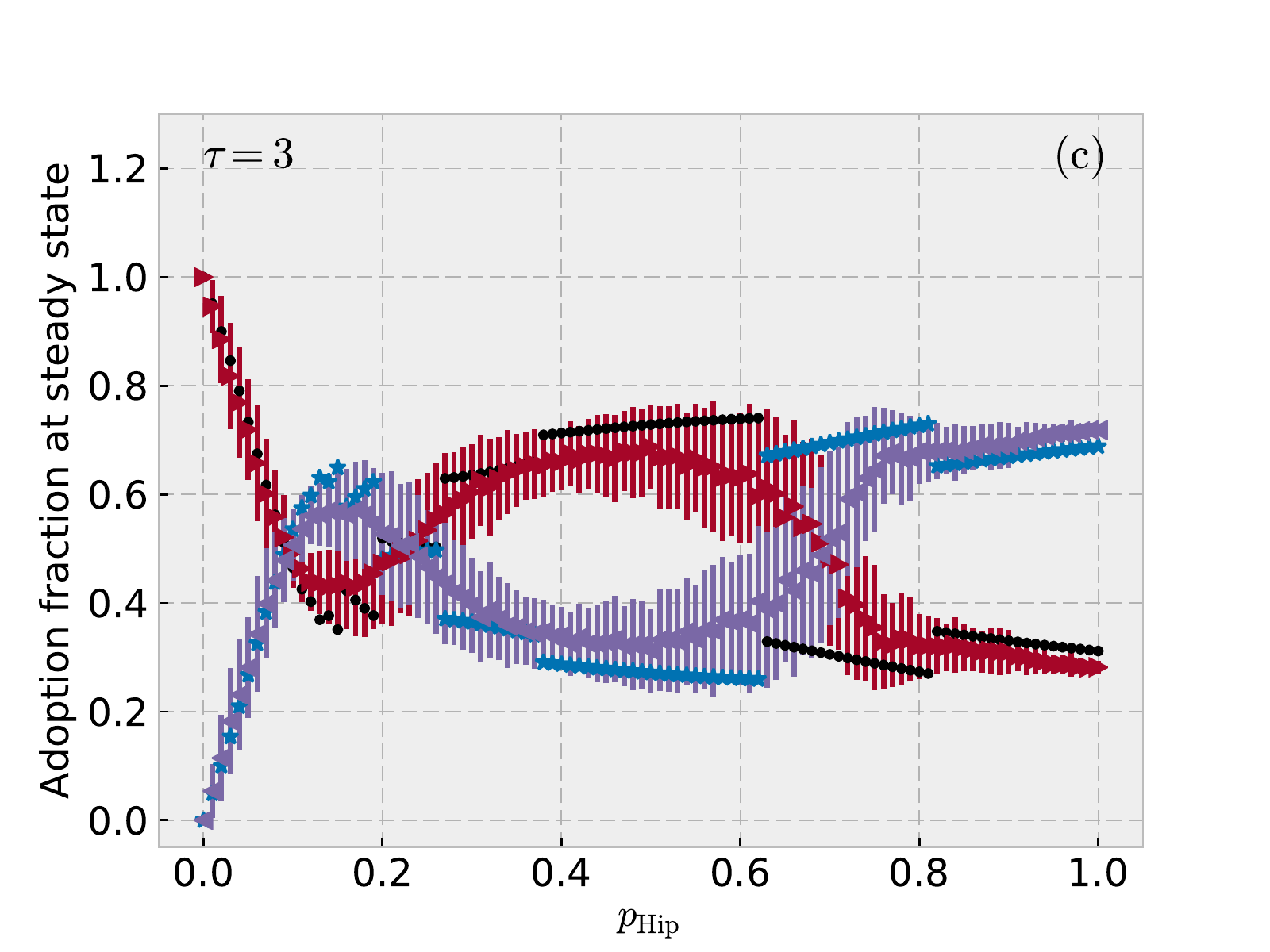}
%\hfill
\includegraphics[width=.43\linewidth]{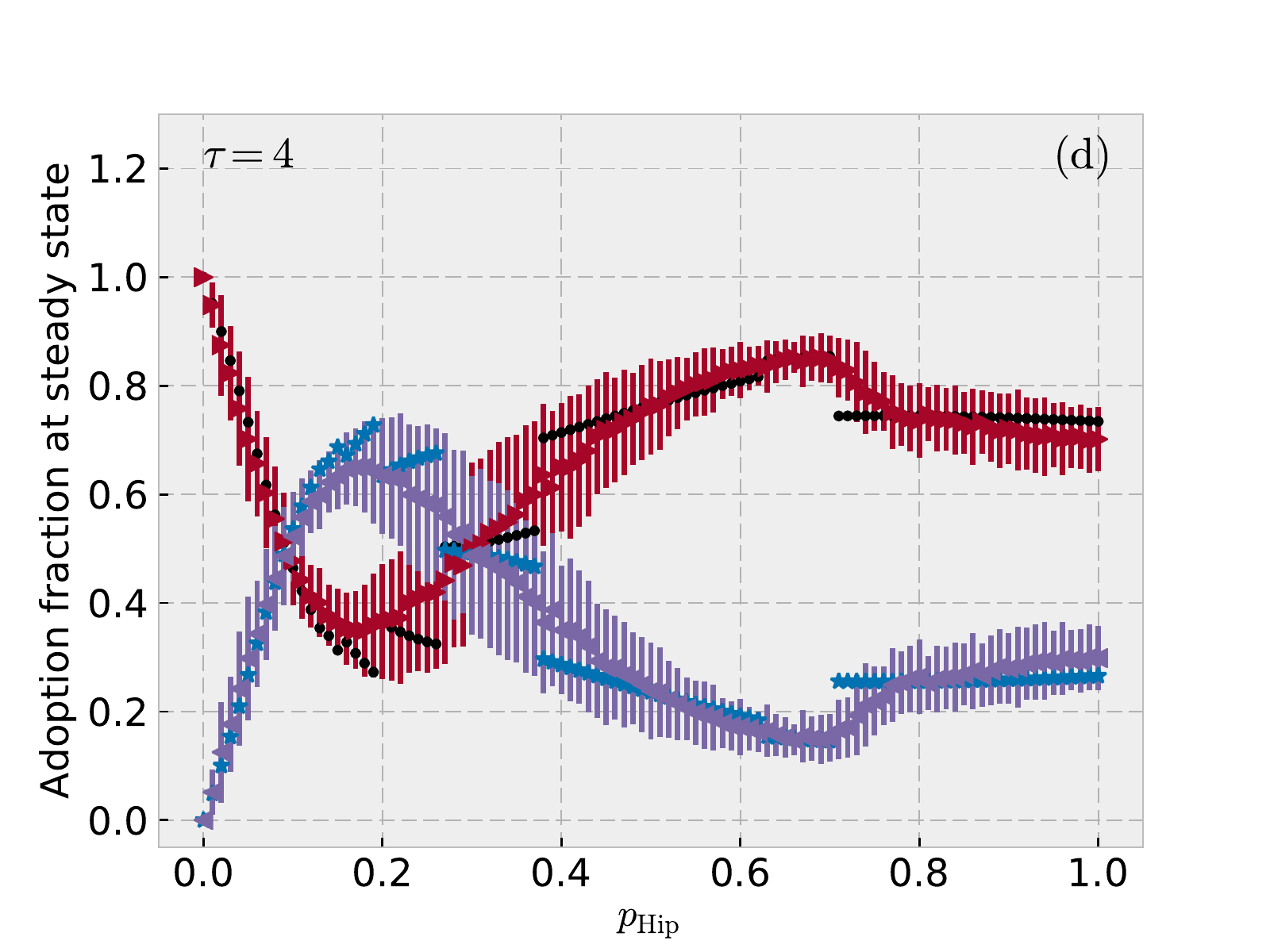}
\\
\includegraphics[width=.43\linewidth]{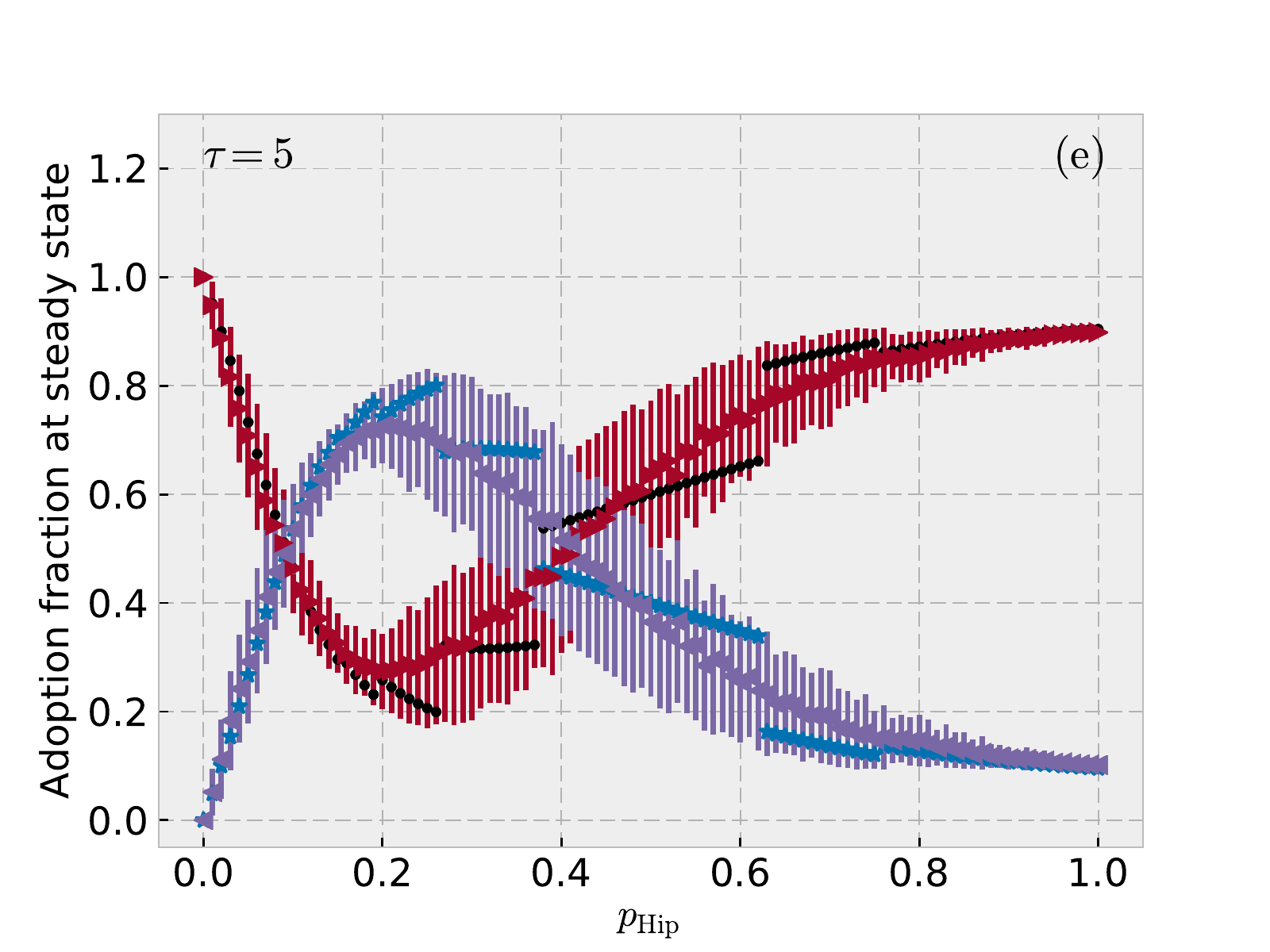}
%\hfill
\includegraphics[width=.43\linewidth]{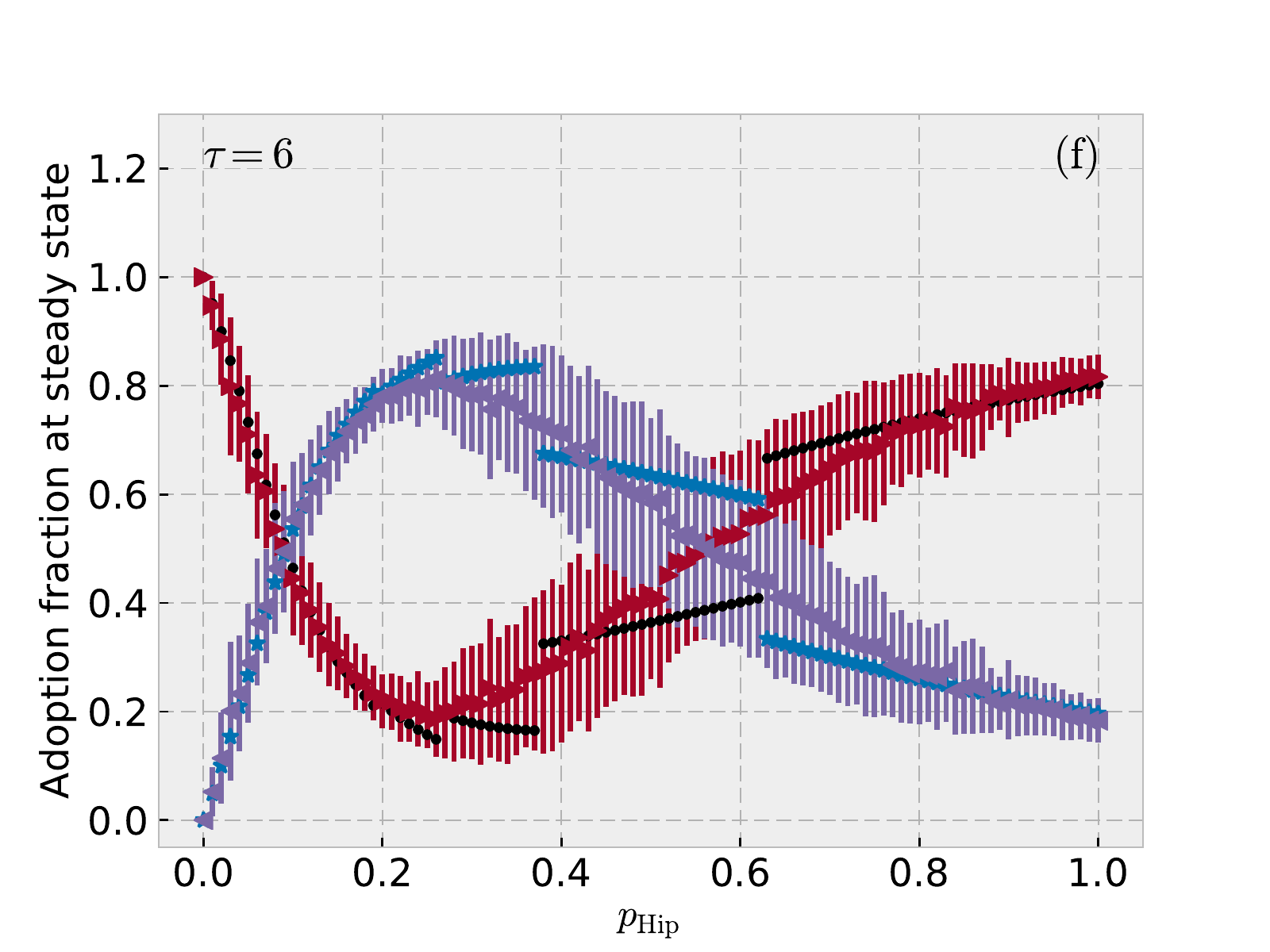}
\caption{Distribution of products at steady state for $10,000$-node $5$-regular configuration-model networks. The different panels give results of simulations of our hipster threshold model with different delay times $\tau$ for the hipster nodes. For each value of $\tau$, we consider hipster probabilities $p_{\mathrm{Hip}} \in [0,1]$ in increments of $0.01$. For each $(p_{\mathrm{Hip}},\tau)$ parameter pair, we simulate the hipster threshold model on $200$ different networks that we construct using a configuration model (in which we connect stubs uniformly at random). The nodes have a threshold of $\phi = 0.19$ with probability $p_0 = 0.8$ and threshold of $\phi = 0.8$ with probability $1-p_0 = 0.2$.
For each simulation, we activate a single node, chosen uniformly at random, with product $A$ at time $t=0$. We stop simulations when products are no longer spreading on the network. We plot the mean fraction of nodes that adopt products $A$ and $B$ in the $200$ realizations and the corresponding standard deviations of the means. (For each $(\tau,p_{\mathrm{Hip}})$ parameter pair, we independently construct $200$ networks, and we also independently determine the initial condition for each network.) For all values of $\tau$, the fraction of nodes that have adopted product $B$ at steady state increases rapidly with $p_{\mathrm{Hip}}$ for small $p_{\mathrm{Hip}}$, reaching $0.5$ at $p_{\mathrm{Hip}} \approx 0.09$. For $\tau = 1$, which we show in panel (a), hipsters have information about the product distribution in the network without any delay, and the steady-state fractions of nodes that adopt products $A$ and $B$ are almost indistinguishable for $p_{\mathrm{Hip}}\gtrapprox 0.09$. For larger values of $\tau$, which we show in panels (b)--(f), the steady-state fractions of nodes that adopt each product varies for $p_{\mathrm{Hip}}\gtrapprox 0.09$. The steady-state fraction of nodes that adopt product $B$ is larger than that for product $A$ for an interval of $p_{\mathrm{Hip}}$ values whose left end is at about $p_{\mathrm{Hip}}\approx 0.09$. For $\tau \in \{ 2,3 \}$ [see panels (b) and (c)], we also observe that more nodes adopt product $B$ for large values of $p_{\mathrm{Hip}}$. The height of the peak in the fraction of product-$B$ adopters above $p_{\mathrm{Hip}}\approx 0.09$ increases with $\tau$, reaching a value above $0.80$ for $\tau = 6$ [see panel (f)]. We also plot the analytically-estimated product-adoption fractions from Eq.~\eqref{eq:analytical_approx}. Our analytical approximation matches the behavior well for small values of $p_{\mathrm{Hip}}$ and large values of $p_{\mathrm{Hip}}$. Between these extremes, however, our approximation has jumps in the steady-state adoption fractions of products; these discontinuities do not arise in our numerical computations. 
}
\label{fig:z_all}
\end{figure*}

We examine our hipster threshold model on the networks for time delays $\tau \in \{1,2,3,4,5,6\}$ and hipster probabilities of $p_{\mathrm{Hip}} \in [0,1]$ (in increments of $0.01$). For each parameter pair $(\tau, p_{\mathrm{Hip}})$, we simulate our model on $200$ different networks. We independently draw the specific sets of networks for different parameter values, so in general they are not the same networks. For each realization, we stop the simulations after the distribution of product adoptions reaches a steady state, and we track the adoption fractions of the two products. From these values, we calculate the mean fraction of nodes that adopt each product over the $200$ realizations and the corresponding standard deviations of the means. We plot these values in Fig.~\ref{fig:z_all}.

For all hipster delay times $\tau$, the steady-state fraction $\rho_{B,{ \mathrm{tot}}}(t\to\infty)$ of nodes that have adopted product $B$ increases rapidly for small $p_{\mathrm{Hip}}$. For $\tau = 1$ [see Fig.~\ref{fig:z_all}(a)], the hipsters have access to information without any delays, and their behavior leads to a balancing of the distributions of products $A$ and $B$. 
If a sufficiently large fraction of the nodes are hipsters, the mean final fraction of nodes that have adopted one product is almost indistinguishable from the other. This occurs for $p_{\mathrm{Hip}} \gtrapprox 0.09$. 

In all examined cases, $\rho_{B,{ \mathrm{tot}}}(t\to\infty)$ approximately equals $\rho_{A,{ \mathrm{tot}}}(t\to\infty)$ (i.e., the steady-state fraction of nodes that have adopted product $A$) for $p_{\mathrm{Hip}} \approx 0.09$. For $\tau \ge 2$ [see Fig.~\ref{fig:z_all}(b)--(f)] and $p_{\mathrm{Hip}} \gtrapprox 0.09$, there exists an interval of $p_{\mathrm{Hip}}$ values in which $\rho_{B,{ \mathrm{tot}}}(t\to\infty) > \rho_{A,{ \mathrm{tot}}}(t\to\infty)$. This interval is larger for larger values of $\tau$, and the peak of $\rho_{B,{ \mathrm{tot}}}(t\to\infty)$ in this interval grows with $\tau$, taking a value above $0.8$ for $\tau = 6$ [see Fig.~\ref{fig:z_all}(f)]. In other words, the fraction of hipsters must be larger than about $0.09$ for product $B$ to become adopted by a larger fraction of the population than product $A$ at steady state.

For $\tau = 2$ [see Fig.~\ref{fig:z_all}(b)], we observe another (and wider) $p_{\rm Hip}$ interval (specifically, at about $[0.35,0.69]$) in which product $B$ beats product $A$. For $p_{ \rm Hip} \gtrapprox 0.69$, product $A$ dominates. Hence, for $\tau = 2$, product $B$ dominates in two $p_{ \rm Hip}$ intervals, and product $A$ dominates in three $p_{\rm Hip}$ intervals. However, $\tau = 3$ [see Fig.~\ref{fig:z_all}(c)] results in two intervals of dominance for each product. Product $B$ is the most popular product at steady state in hipster-probability intervals starting at $p_{\mathrm{Hip}} \approx 0.09$ and $p_{\mathrm{Hip}} \approx 0.69$. Our simulations with $\tau \geq 4$ result in a single $p_{\mathrm{Hip}}$ interval in which product $B$ is more popular than product $A$ at steady state. From the standard deviations, we see that different realizations with the same parameter values can yield rather different results.

Our analytical approximation and numerical computations match well for small and very large $p_{\mathrm{Hip}}$. However, our approximation includes jumps in the fraction of adopters for each of the products, and we do not observe such discontinuities in our simulations. The mean fraction of nodes that adopt a product in the discarded realizations is $0.0001$, which is much less than the threshold of $0.10$.

%%%%

\subsection{$3$-regular configuration-model networks} \label{three}

\begin{figure*}[th!]
\centering
\includegraphics[width=.43\linewidth]{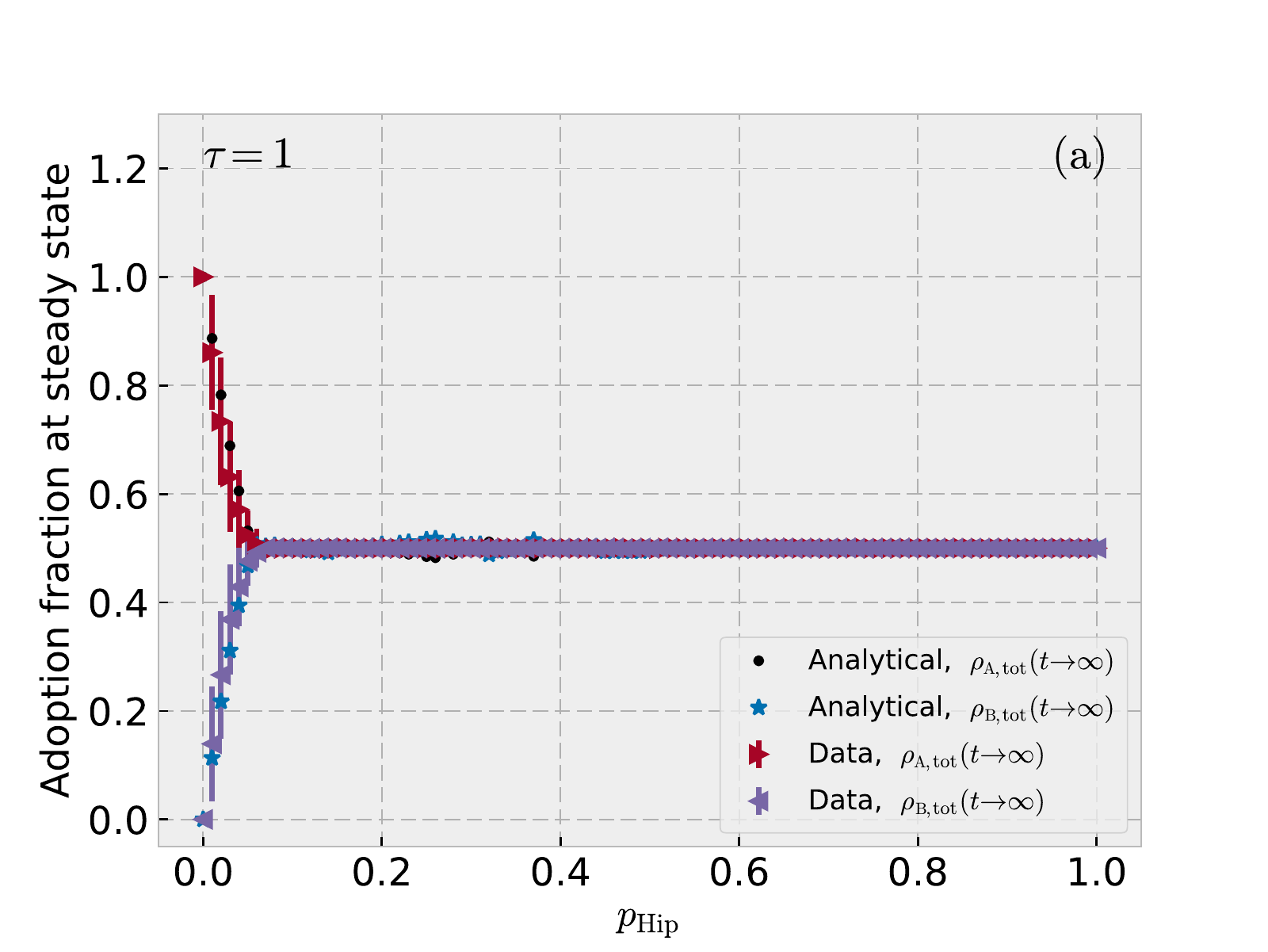}
%\hfill
\includegraphics[width=.43\linewidth]{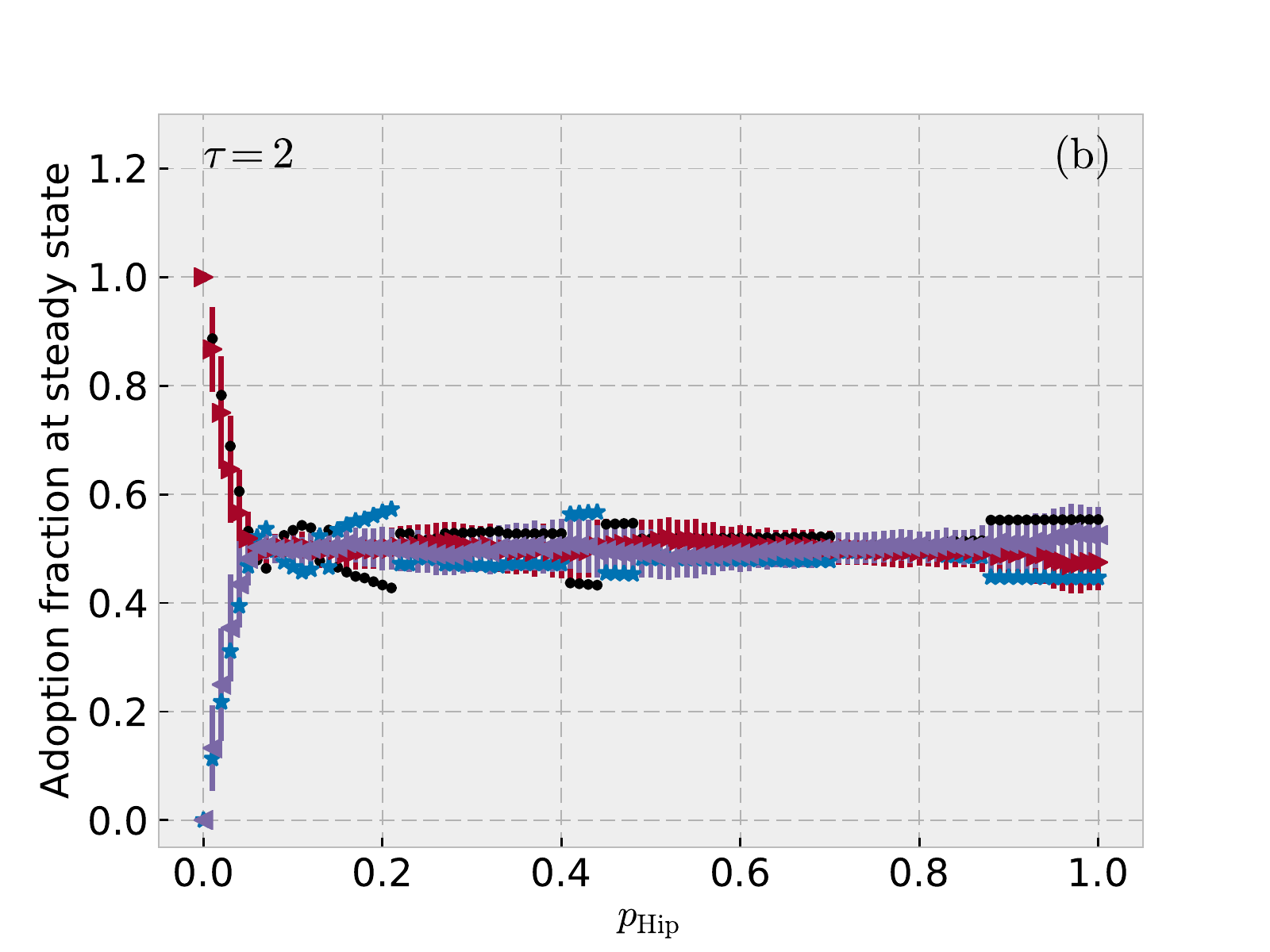}
\\
\includegraphics[width=.43\linewidth]{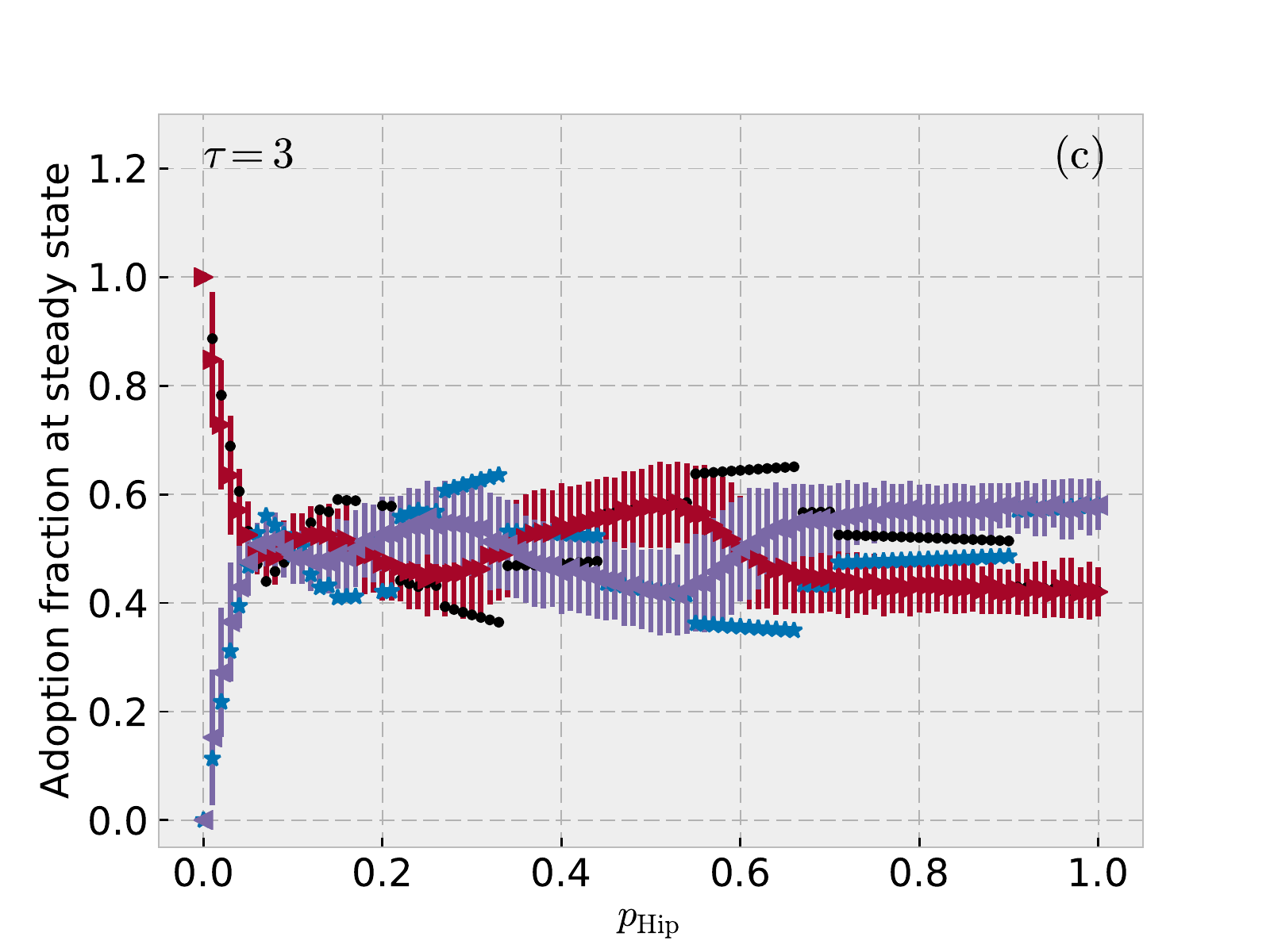}
%\hfill
\includegraphics[width=.43\linewidth]{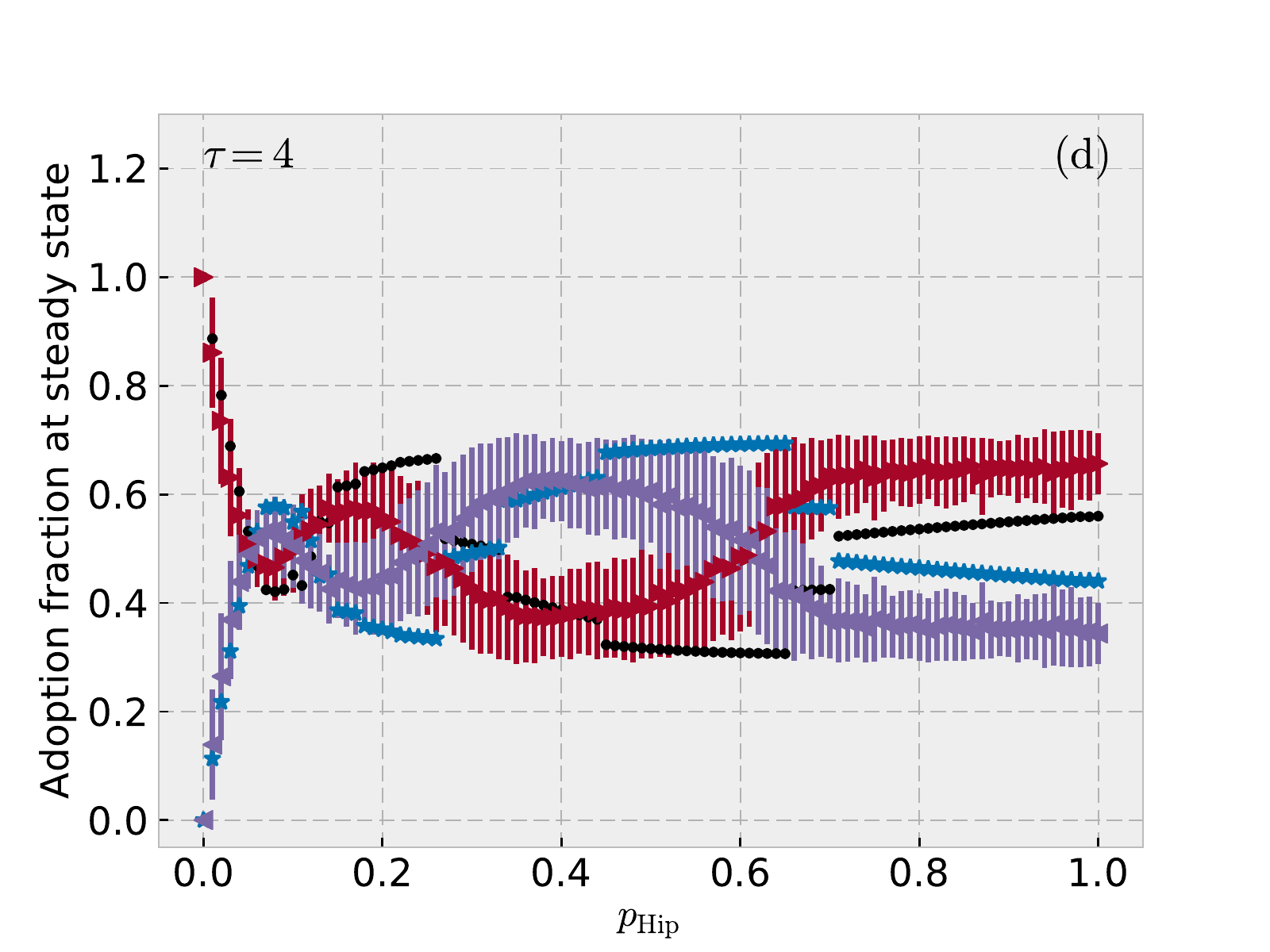}
\\
\includegraphics[width=.43\linewidth]{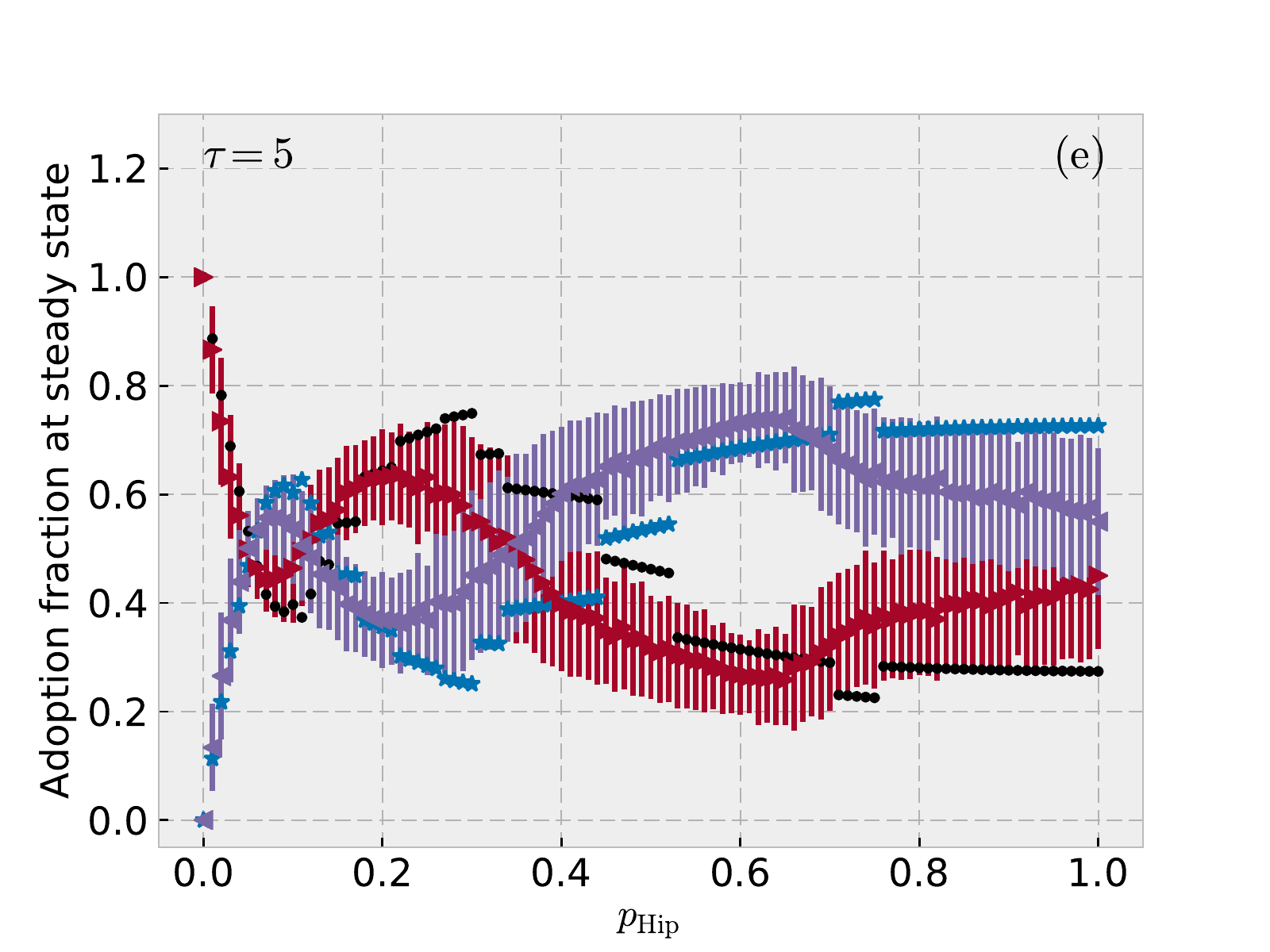}
%\hfill
\includegraphics[width=.43\linewidth]{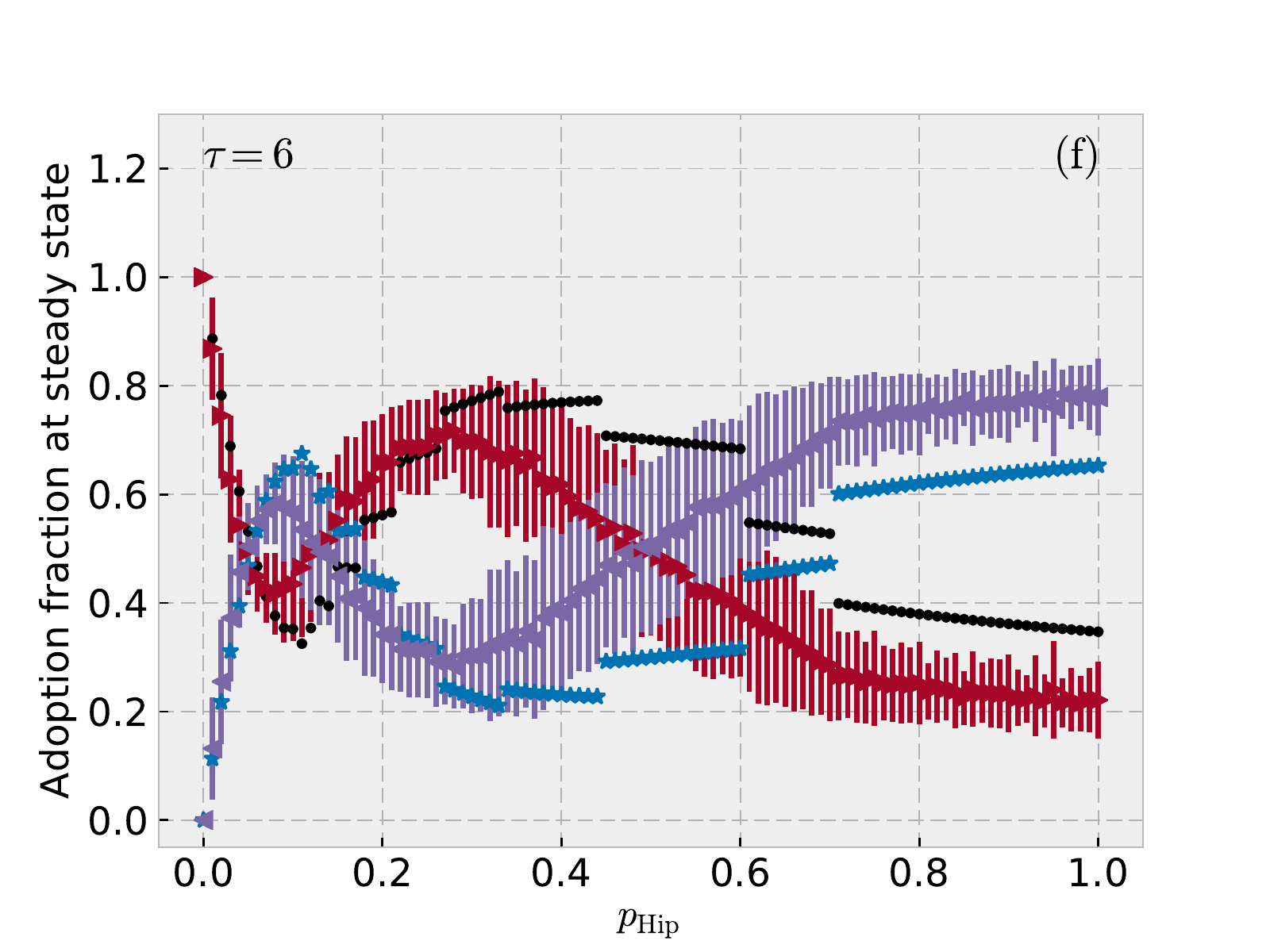}
\caption{Distribution of products at steady state for $10,000$-node $3$-regular configuration-model networks. The different panels give results of simulations of our hipster threshold model with different delay times $\tau$ for the hipster nodes. For each value of $\tau$, we consider hipster probabilities $p_{\mathrm{Hip}} \in [0,1]$ in increments of $0.01$. For each $(p_{\mathrm{Hip}},\tau)$ parameter pair, we simulate the model on $200$ different networks that we constructed using a configuration model (in which we connect stubs uniformly at random). The nodes have a threshold of $\phi = 0.3$ with probability $p_0 = 0.8$ and threshold of $\phi = 0.65$ with probability $1-p_0 = 0.2$. For each simulation, we activate a single node, chosen uniformly at random, with product $A$ at time $t=0$. We stop our simulations when the products are no longer spreading on the network.  We plot the mean steady-state fraction of nodes that adopt products $A$ and $B$ in the $200$ realizations and the corresponding standard deviations of the means. (For each $(\tau,p_{\mathrm{Hip}})$ parameter pair, we independently construct $200$ networks, and we also independently determine the initial condition for each network.) For all values of $\tau$, the steady-state fraction of nodes that adopt product $B$ increases rapidly with $p_{\mathrm{Hip}}$ for small $p_{\mathrm{Hip}}$, reaching $0.5$ at $p_{\mathrm{Hip}} \approx 0.06$. For $\tau = 1$, which we show in panel (a), hipsters have information about the product distribution in the network without any delay, and the steady-state fractions of nodes that adopt products $A$ and $B$ are almost indistinguishable for $p_{\mathrm{Hip}}\gtrapprox 0.06$. 
For $\tau = 2$, which we show in (b), the fractions of nodes that adopt the two products are similar (though one can see some interesting dynamics) until $p_{\mathrm{Hip}} \approx 0.93$, above which product $B$ is the more-popular product. For larger values of $\tau$ [see panels (c)--(f)], the fraction of nodes that adopt each product varies for $p_{\mathrm{Hip}}\gtrapprox 0.05$. The height of the peak, which occurs at $p_{\mathrm{Hip}} \approx 0.08$, of the node fraction that adopt product $B$ increases with $\tau$, reaching a value of over $0.6$ for $\tau = 6$ [see panel (f)]. For all time delays $\tau \geq 2$ [see panels (b)--(f)], the maximum steady-state fraction that adopts product $B$ does not take place at $p_{\mathrm{Hip}}$ values near $0.08$; instead, it occurs for much larger values of $p_{\mathrm{Hip}}$. We also plot our analytically-estimated fractions of product adoption from Eq.~\eqref{eq:analytical_approx}. Our approximation matches well with our computations well for small values of $p_{\mathrm{Hip}}$. For $p_{\mathrm{Hip}} \gtrapprox 0.06$, however, our  approximation does not do well. 
Our analytical solution includes jumps in the steady-state adoption fractions of the products, but these do not arise in our numerical simulations.}\label{fig:z3}
\end{figure*}

We now examine our hipster threshold model on $3$-regular configuration-model networks. Suppose that a fraction $p_0 = 0.8$ of the nodes have a threshold of $\phi = 0.3$ and that the remaining fraction $1 - p_0 = 0.2$ of the nodes have a threshold of $\phi = 0.65$. We perform simulations as in the $5$-regular configuration-model networks (see Section \ref{five}) and show our results in Fig.~\ref{fig:z3}. 

Our results on $3$-regular configuration-model networks differ from those on $5$-regular configuration-model networks in several ways. One interesting result is that the fractions that adopt products $A$ and $B$ are very similar for $\tau = 2$ [see Fig.~\ref{fig:z3}(b)] and $p_{\mathrm{Hip}} \in [0.06, 0.93]$. Additionally, for all examined $\tau \ge 3$ [see Fig.~\ref{fig:z3}(c)--(f)], the $\rho_{B,{ \mathrm{tot}}}(t\to\infty)$ curve on $3$-regular networks has one more maximum than the corresponding curve on the $5$-regular configuration-model networks. 

\begin{figure}[ht!]
\centering
\includegraphics[width=\linewidth]{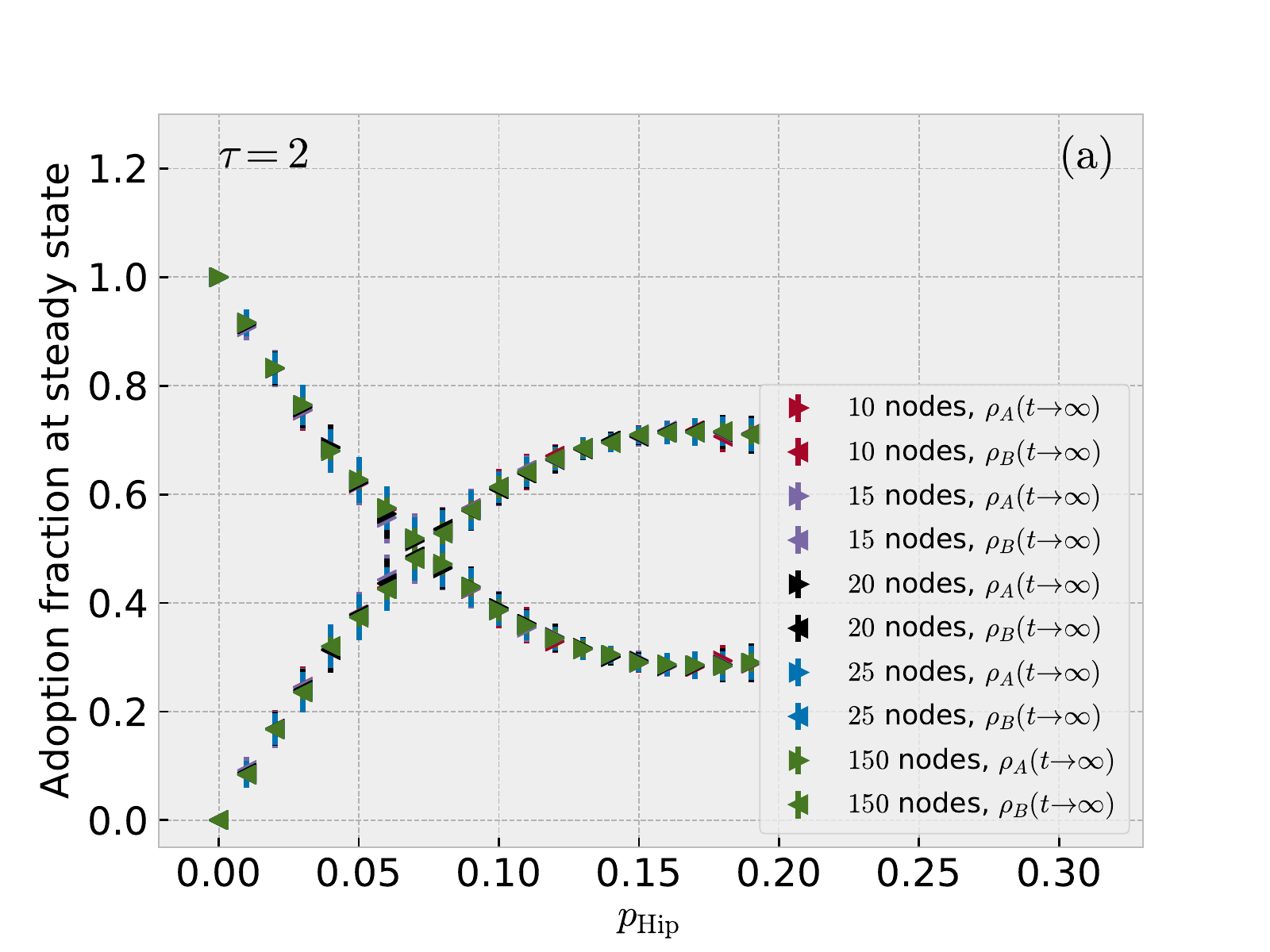}
\includegraphics[width=\linewidth]{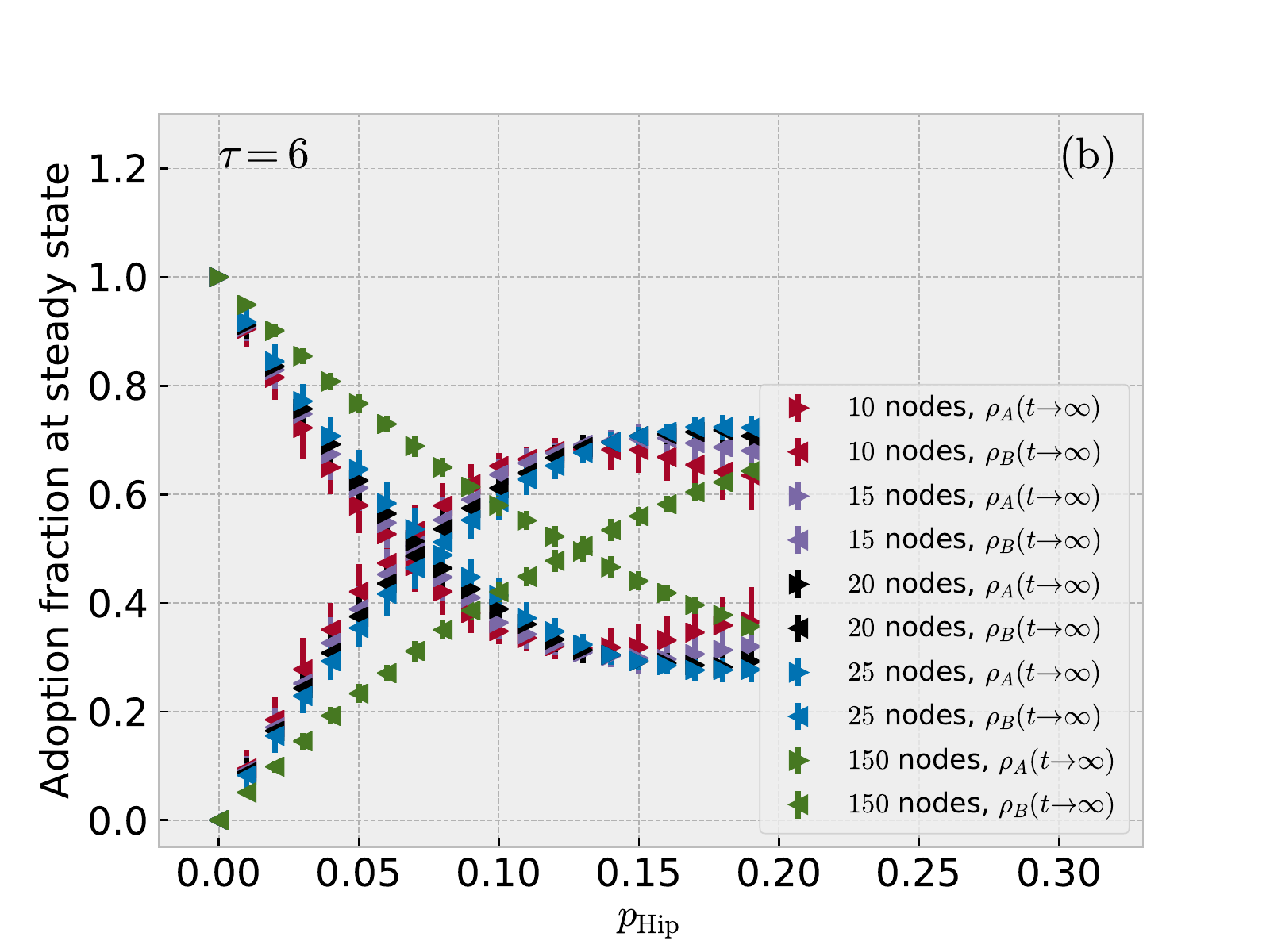}
\caption{Distribution of products at steady state for $10,000$-node $3$-regular configuration-model networks with different seed sizes, where all seed nodes adopt product $A$. We show the distribution of products for delays of (a) $\tau = 2$ and (b) $\tau = 6$. With $\tau = 2$, the adoption fraction is indistinguishable for the different seed sizes. For $\tau = 6$, our results vary for different seed sizes, but the qualitative behavior is consistent across all cases: the steady-state fraction of nodes that adopt product $B$ increases rapidly with $p_{\rm Hip}$, and equal fractions adopt products $A$ and $B$ at a value of $p_{\rm Hip}$ that increases slowly with seed size. For seed sets with $10$ to $25$ nodes, equal fractions of nodes adopt the two products at $p_{\rm Hip}\approx 0.07$; for a seed set $150$ nodes, equal fractions adopt the two products at $p_{\rm Hip}\approx 0.13$.
}
\label{fig:seedsize}
\end{figure}

On $3$-regular configuration-model networks with $\tau \geq 2$, we observe that $\rho_{B,{ \mathrm{tot}}}(t\to\infty)$ first becomes larger than $\rho_{A,{ \mathrm{tot}}}(t\to\infty)$ at about $p_{\mathrm{Hip}}\approx 0.06$, which is lower than the hipster probability that we observed for the analogous result for $5$-regular configuration-model networks. As we show in Fig.~\ref{fig:seedsize}, this transition may or may not change with seed size, depending on the value of the delay $\tau$. For $\tau=2$, the transition occurs at the same probability when we seed more nodes with product $A$; however, for the larger delay value $\tau = 6$, the transition moves towards larger probabilities for progressively larger sets of seed nodes who have adopted product $A$. When the sed size is $1$, we observe in Fig.~\ref{fig:z3} that the height of the first peak is lower when we simulate our model on $3$-regular configuration-model networks than was the case for $5$-regular configuration-model networks. For large $p_{\mathrm{Hip}}$, the most popular product at steady state is the same for $\tau = \{3,4\}$ [see Fig.~\ref{fig:z3}(c,d)] as for the same delay times on the $5$-regular configuration-model networks, while it is opposite to that on the $5$-regular networks for $\tau = \{5,6 \}$ [see Fig.~\ref{fig:z3}(e,f)]. For many parameter pairs, the standard deviations of the outcomes are large, indicating that realizations with identical parameters can yield very different outcomes.

Our analytical approximation and numerical simulations match well for $p_{\mathrm{Hip}}\lessapprox 0.05$. However, for values of $p_{\mathrm{Hip}}$ that are larger than about $0.05$, our analytical approximation again has jumps that we do not observe in computations. Our analytical approximation also does not predict the fraction of nodes that adopt each product for $p_{\mathrm{Hip}}=1$ as well as it did on $5$-regular configuration-model networks. This may be because $3$-regular configuration-model networks have a higher edge density than the $5$-regular configuration-model networks, so the former depart rather significantly from satisfying a local-tree hypothesis (on which our analytical approximation relies). The mean fraction of nodes that were activated in the discarded realizations was $0.0002$, which is much less than the threshold of $0.10$.

%%%%%

\subsection{\ER {} networks}

We now examine our hipster threshold model on \ER { }(ER) networks. Specifically, we examine $G(N,p)$ graphs, in which one specifies the total number $N$ of nodes, and each pair of nodes is linked independently with probability $p$. We choose the expected mean degree of the networks to be $z=5$ (so the probability of an edge between any two nodes is $p=z/N$) to match the mean degree of the $5$-regular configuration-model networks that we examined in Section \ref{five}. 

\begin{figure*}[ht!]
\centering
\includegraphics[width=.43\linewidth]{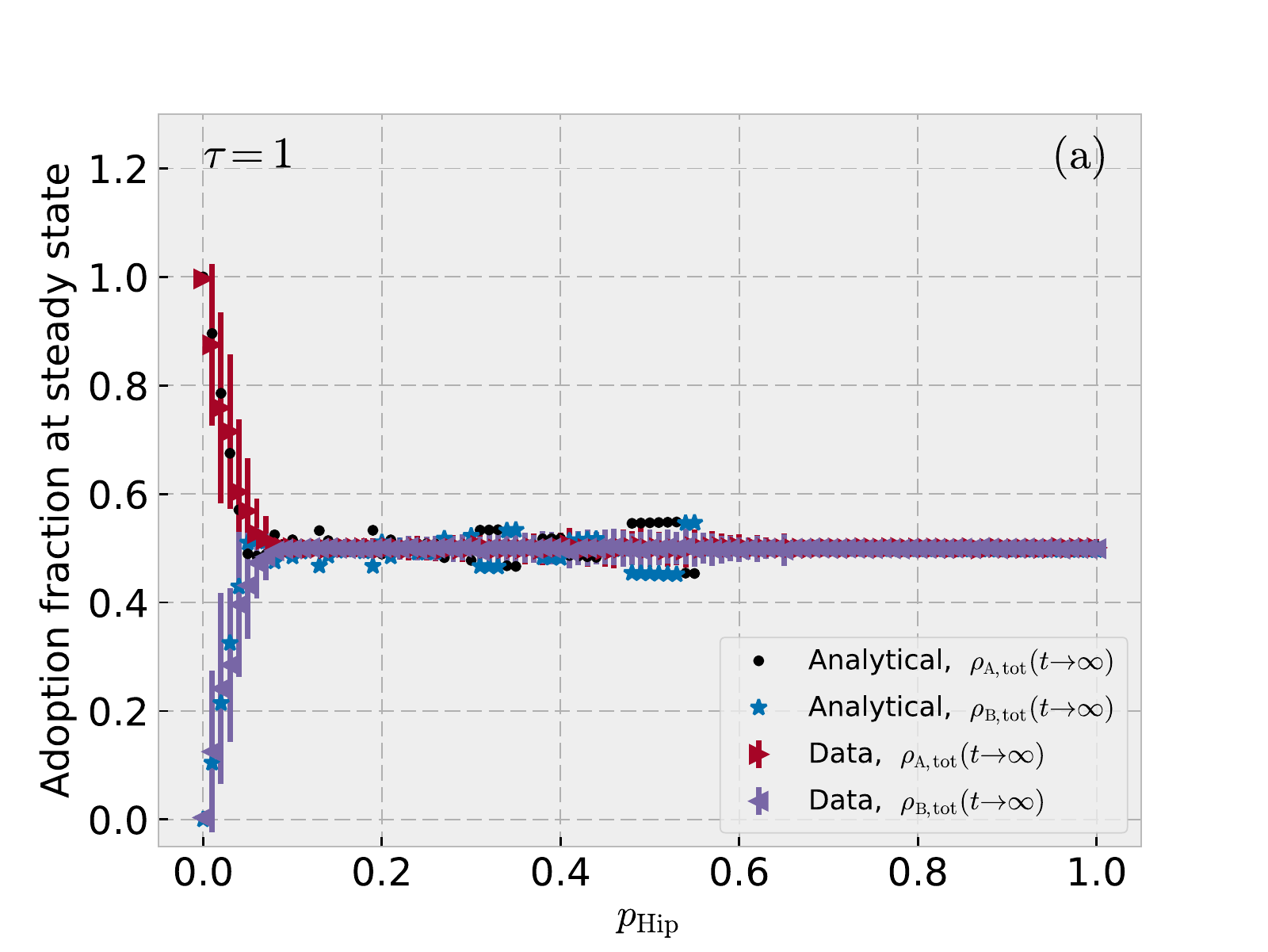}
%\hfill
\includegraphics[width=.43\linewidth]{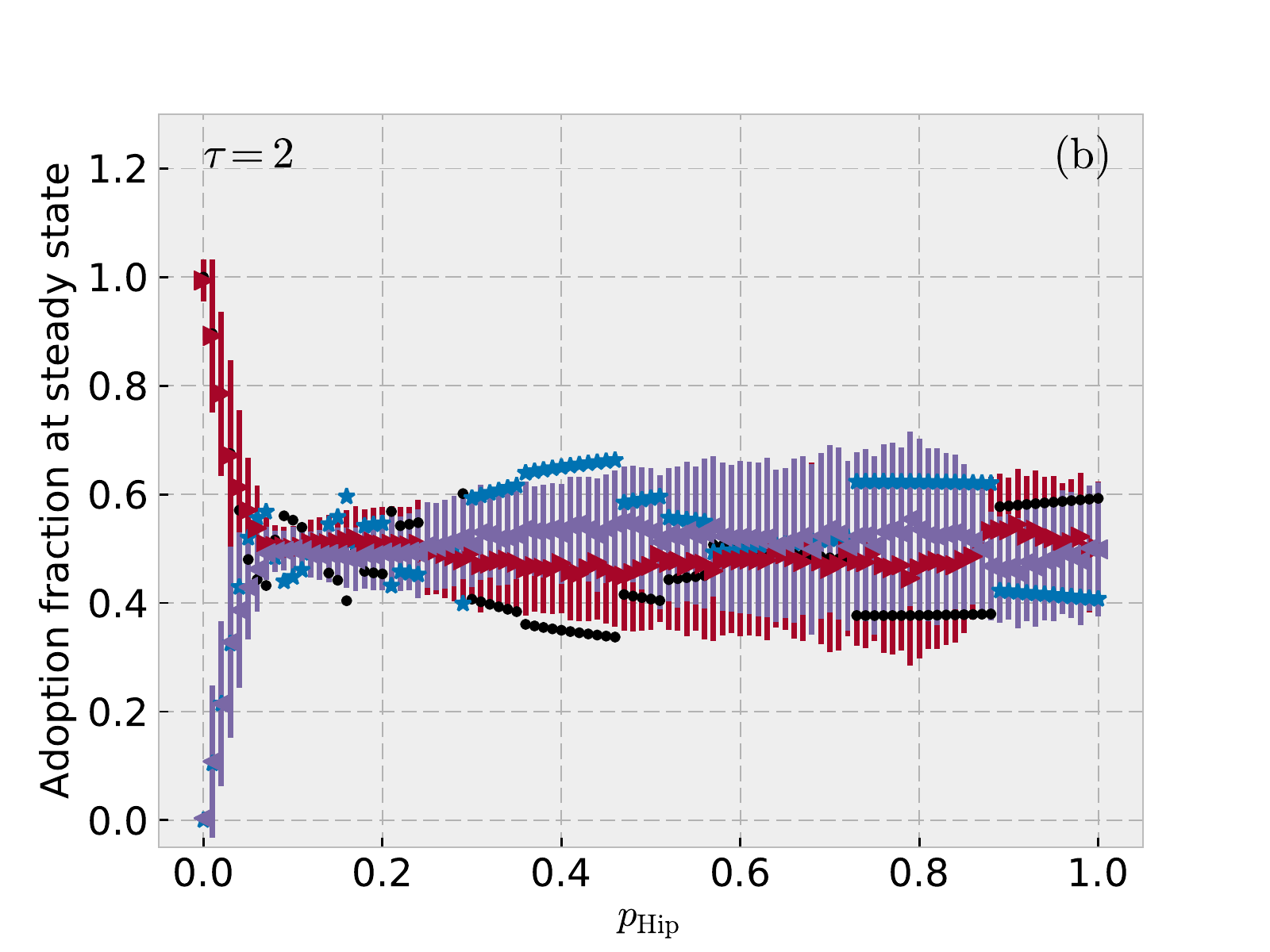}
\\
\includegraphics[width=.43\linewidth]{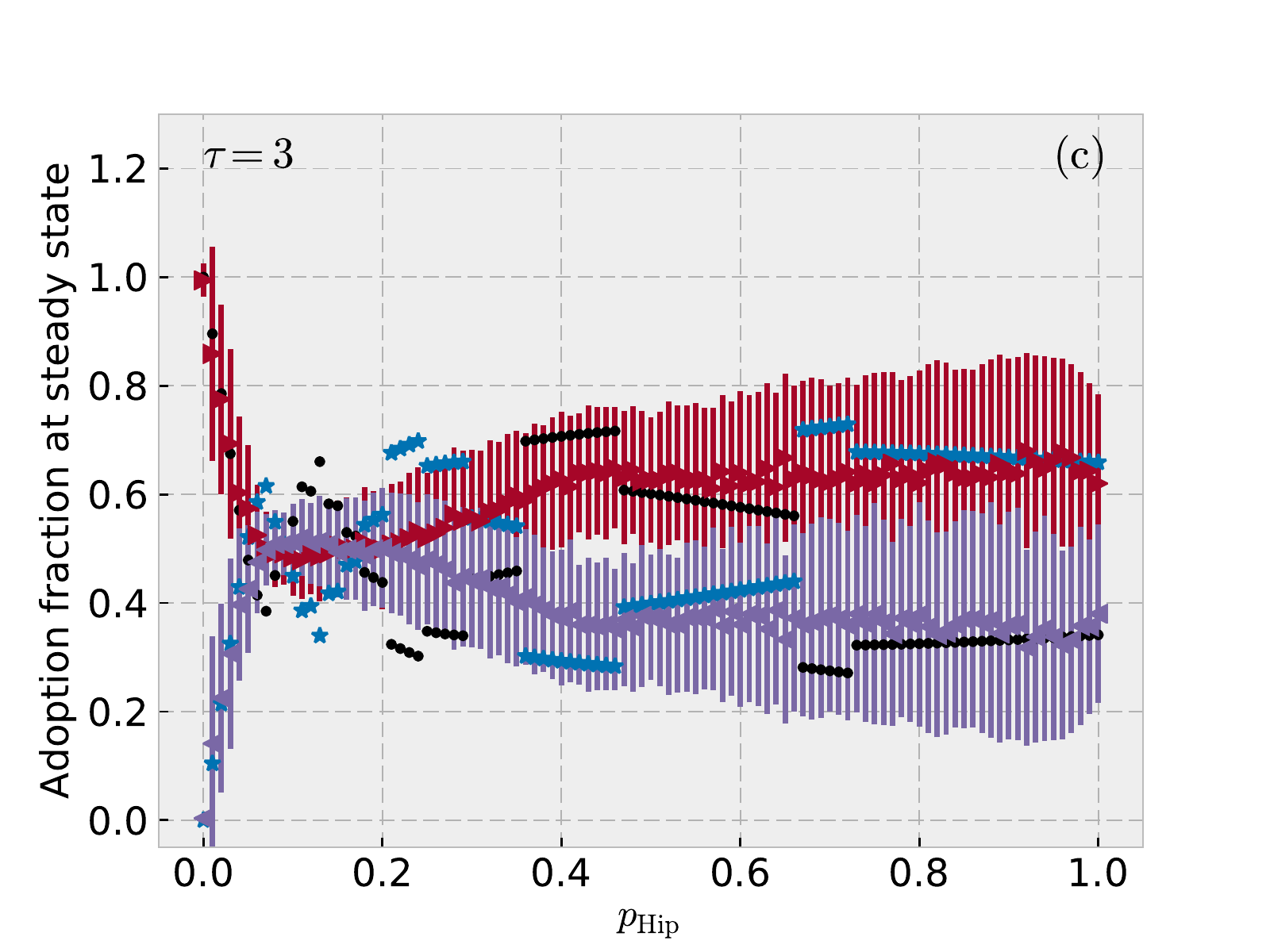}
%\hfill
\includegraphics[width=.43\linewidth]{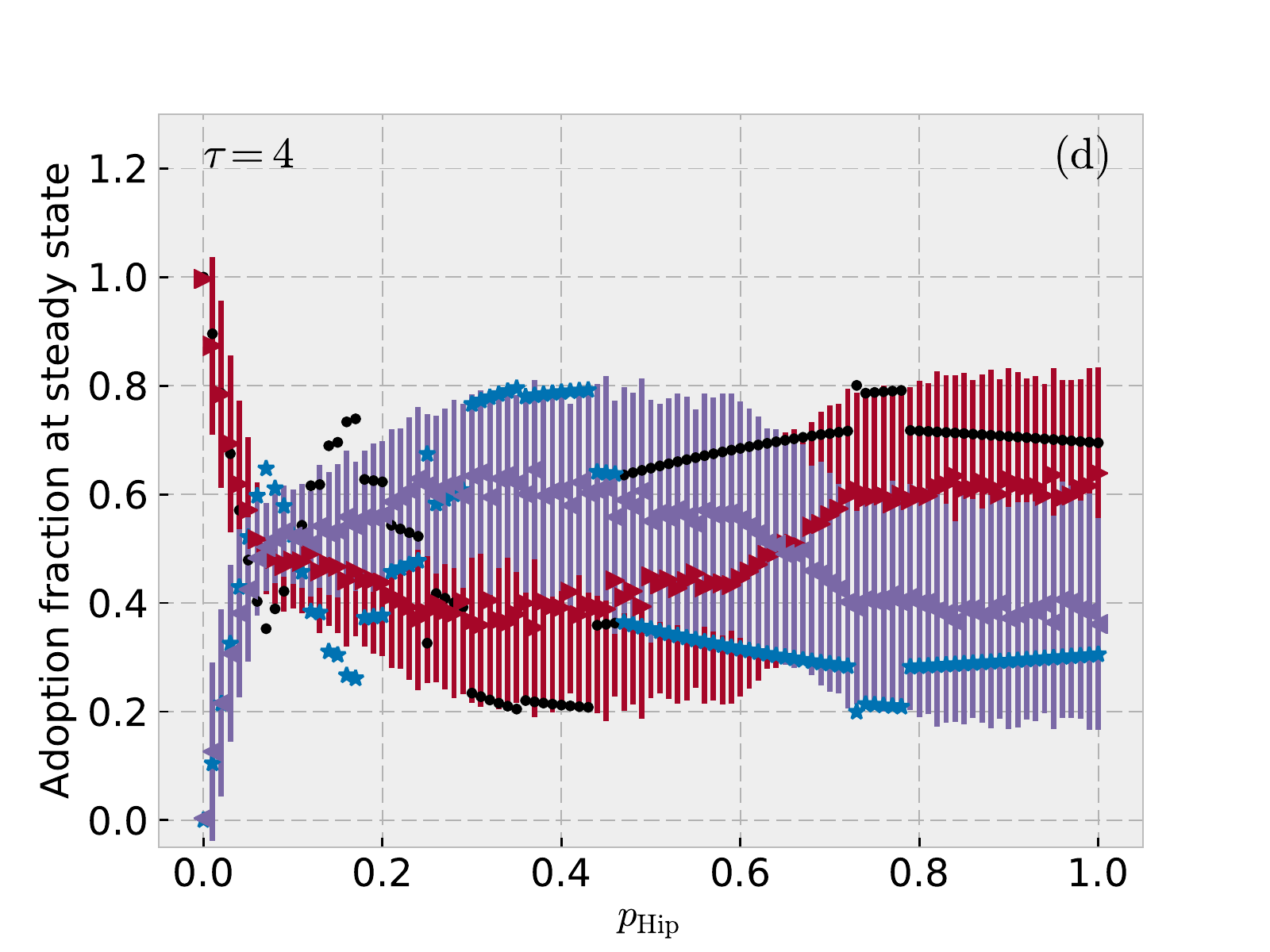}
\\
\includegraphics[width=.43\linewidth]{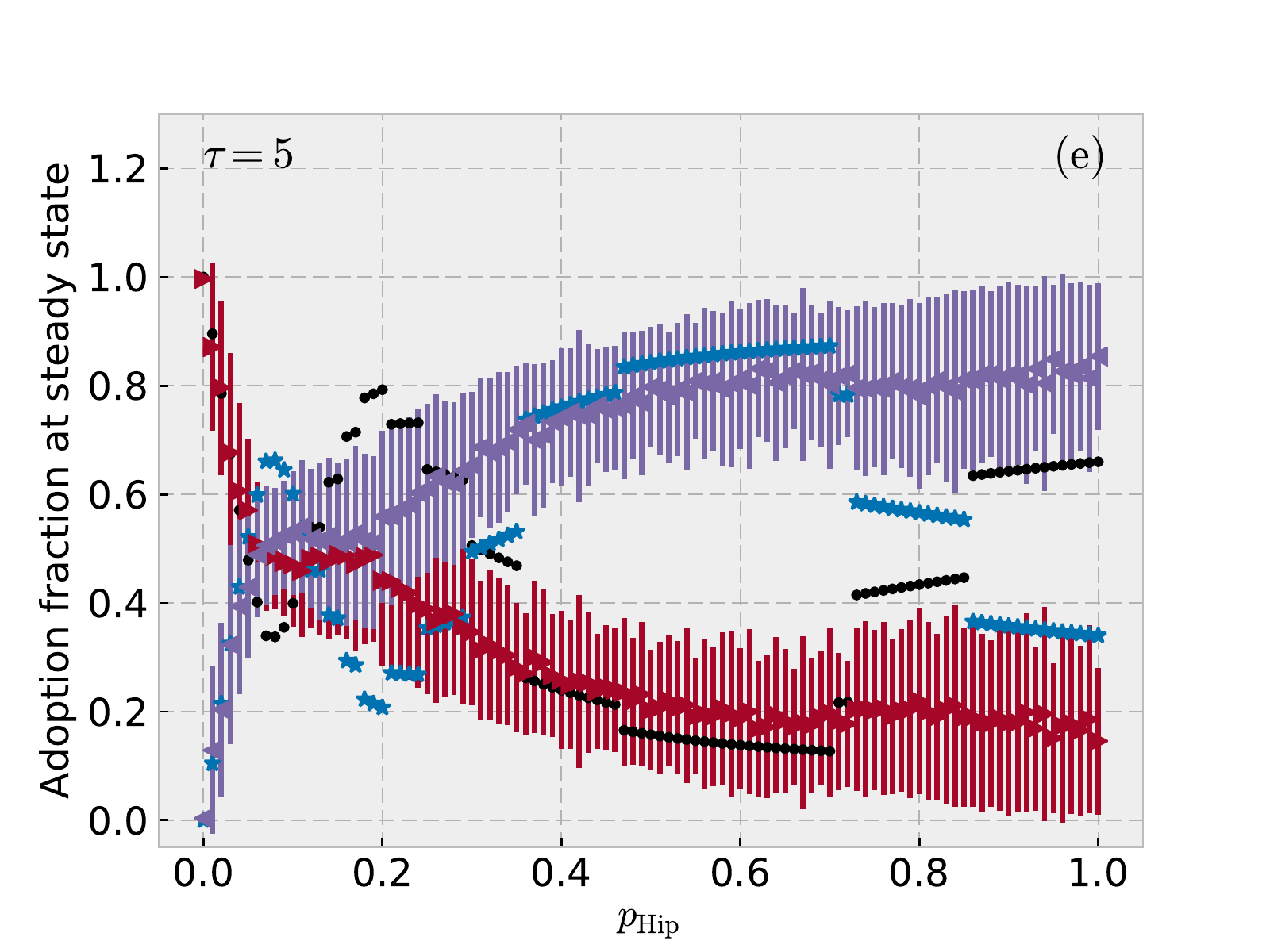}
%\hfill
\includegraphics[width=.43\linewidth]{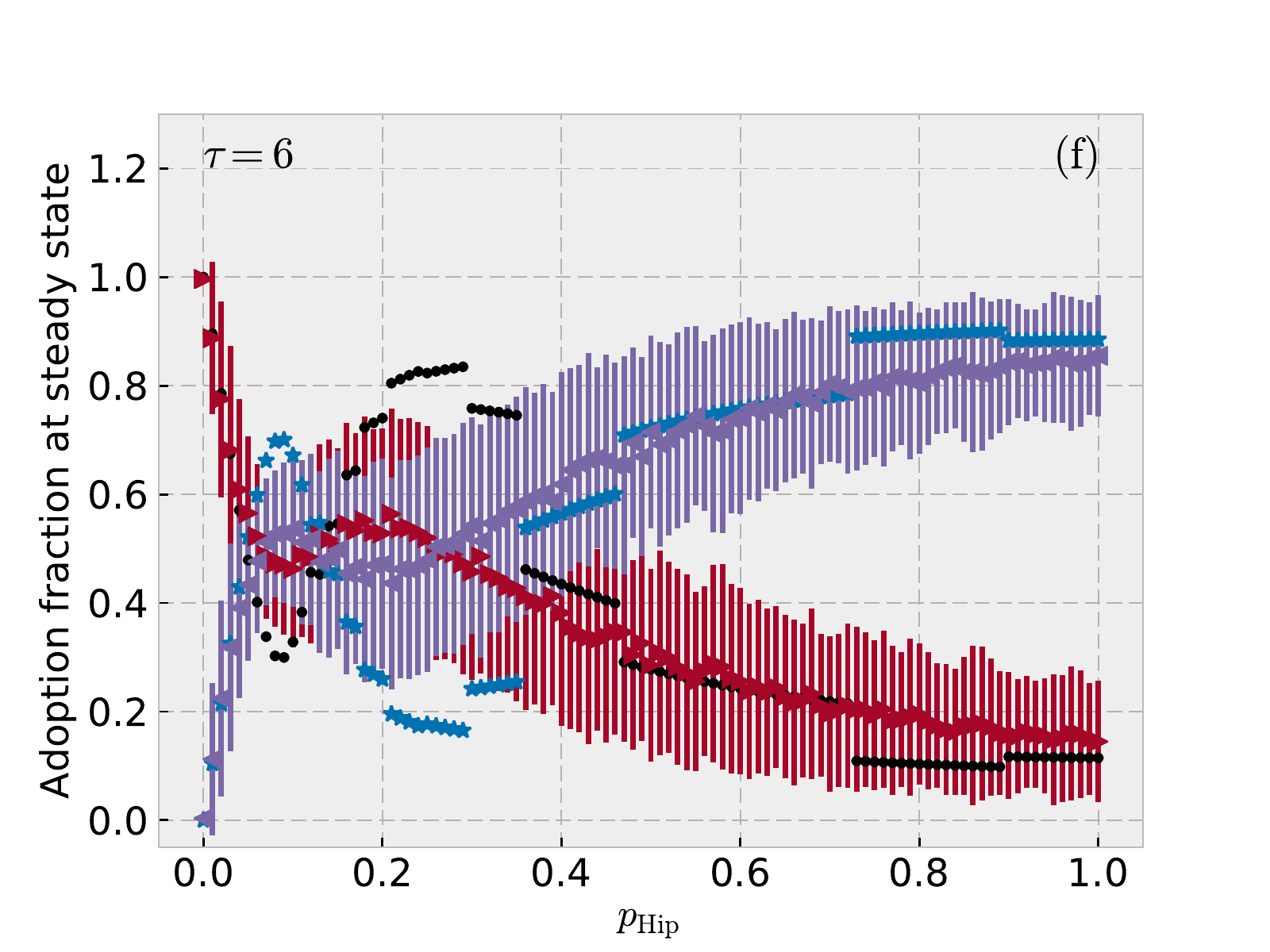}

\caption{Distribution of products at steady state for $10,000$-node \ER { } networks with an expected mean degree of $z=5$. The different panels give results of simulations of our hipster threshold model with different delay times $\tau$ for the hipster nodes. For each value of $\tau$, we consider hipster probabilities $p_{\mathrm{Hip}} \in [0,1]$ in increments of $0.01$. For each $(\tau, p_{\mathrm{Hip}})$ parameter pair, we simulate the model on $200$ different networks and initial conditions. Each node has a threshold of $\phi = 0.2$. For each simulation, we activate a single node, chosen uniformly at random, with product $A$ at time $t=0$. We stop simulations when products are no longer spreading on the network. We plot the mean steady-state fraction of nodes that adopt products $A$ and $B$ in the $200$ realizations and the corresponding standard deviations of the means.   (For each $(\tau,p_{\mathrm{Hip}})$ parameter pair, we independently construct $200$ networks, and we also independently determine the initial condition for each network.) For all values of $\tau$, the fraction of nodes that adopt product $B$ increases rapidly with $p_{\mathrm{Hip}}$ for small $p_{\mathrm{Hip}}$, reaching $0.5$ at $p_{\mathrm{Hip}} \approx 0.07$.  For $\tau = 1$, which we show in panel (a), hipsters have information about the product distribution in the network without any delay, and the steady-state fractions of nodes that adopt products $A$ and $B$ are almost indistinguishable for $p_{\mathrm{Hip}} \gtrapprox 0.07$. For larger values of $\tau$ [see panels (b)--(f)], the fraction of nodes that adopt each product varies for $p_{\mathrm{Hip}} \gtrapprox 0.07$. For all $\tau \ge 3$ [see panels (c)--(f)] the fraction of nodes that adopt product $B$ is largest for a small interval of $p_{\mathrm{Hip}}$ around $p_{\mathrm{Hip}} \approx 0.10$. For $\tau\ge 4$ [see panels (d)--(f)], we observe an additional, large-$p_{\mathrm{Hip}}$ interval in which a majority of the nodes adopt product $B$.
We also plot the analytically-estimated fractions of product adoption from Eq.~\eqref{eq:analytical_approx}. The analytical curves and numerical simulations match well for small values of $p_{\mathrm{Hip}}$. For larger hipster probabilities, however, our analytical approximation is not accurate. For $\tau = 5$ [see panel (e)], it predicts incorrectly that product $A$ is the more-popular product at steady state for large values of $p_{\mathrm{Hip}}$. Our analytical results also include jumps in the steady-state adoption fractions of products that are not present in our numerical simulations.}
\label{fig:ER_all}
\end{figure*}

We assign the same threshold $\phi_i = \phi^* = 0.2$ to each node. With this threshold, all nodes with degree $k\le 5$ are vulnerable. We again consider $p_{\mathrm{Hip}} \in [0,1]$ (in increments of $0.01$) and $\tau \in \{1,2,3,4,5,6\}$.
 For each parameter pair $(\tau, p_{\mathrm{Hip}})$, we simulate the dynamics on $200$ different networks, stop the simulations after reaching a steady state, track the final fractions of nodes that have adopted each of the products, and calculate the corresponding mean and standard deviation of the mean from these data. As in prior simulations, we use a different set of 200 networks for each parameter value. We plot our results in Fig.~\ref{fig:ER_all}.

As with our simulations on $5$-regular and $3$-regular configuration-model networks, the absence of time delay (i.e., $\tau=1$) in the information possessed by hipsters results in $\rho_{B,{ \mathrm{tot}}}(t\to\infty)$ being almost indistinguishable from $\rho_{A,{ \mathrm{tot}}}(t\to\infty)$ [see Fig.~\ref{fig:ER_all}(a)]. For all examined values of $\tau$, we observe that $\rho_{B,{ \mathrm{tot}}}(t\to\infty)$ again increases rapidly for small values of $p_{\mathrm{Hip}}$. For $p_{\mathrm{Hip}} \approx 0.07$, we observe that $\rho_{A,{ \mathrm{tot}}}(t\to\infty)$ and $\rho_{B,{ \mathrm{tot}}}(t\to\infty)$ have similar steady-state fractions, although one can also observe rather interesting dynamics. For large values of $p_{\mathrm{Hip}}$, the same product becomes the more-popular one at steady state as with the $3$-regular configuration-model networks for delay times $\tau = \{4,5,6\}$ [see Fig.~\ref{fig:ER_all}(d)--(f)], but product $A$ is the more-popular one for $\tau = 3$ [see Fig.~\ref{fig:ER_all}(c)]. We generally observe large standard deviations of the outcomes of realizations with given parameter values. The mean fraction of nodes that adopt a product in the discarded realizations is $0.0072$. This is larger than what we observed for $3$-regular and $5$-regular configuration-model networks, but it is still much smaller than the threshold of $0.10$. 

Our analytical approximation and numerical simulations once again match well for small values of $p_{\rm Hip}$ (specifically, for $p_{\rm Hip} \lessapprox 0.07$). For larger values of $p_{{\rm Hip}}$, our analytical approximation has jumps in the fraction that adopt each product, which we again do not observe in the simulations. One possibility, which we suggested in our discussion of 3-regular configuration-model networks in Section \ref{three}, is whether our analytical approximation is running into problems because we are considering networks that are not locally tree-like (although similar approximations are known to be effective for many networks that are not locally tree-like \cite{localtreeapprox_mason}). Additionally, note that the mean local clustering coefficient for our ER networks with $z = 5$ is $0.00058 \pm 0.00018$, so our networks have very few $3$-cycles. If we ignore which product is adopted and pretend that the two products are the same, we recover the usual WTM model; the present paper uses an analytical approximation that is known to work in that situation \cite{gleeson2013PRX}. Our own recent work has demonstrated that this type of analytical approximation is also effective for a WTM augmented with ``synergistic'' social influence from nodes other than nearest neighbors \cite{juul2017synergistic}, so the incorporation of different nodes (rather than the lack of a locally tree-like network structure) appears to be the likely cause of the breakdown of the approximation, especially given that the approximation becomes worse as we increase the hipster probability $p_{\rm Hip}$.

%%%%%

\subsection{The {\sc Northwestern25} Facebook network}

In Section \ref{three}, we showed simulations of our hipster threshold model model on the {\sc Northwestern25} network from the {\sc Facebook100} data set for two choices of the $(p_{Hip}, \tau)$ parameter pair. We now examine the model on the {\sc Northwestern25} network more systematically by considering more initial conditions and a wider variety of parameter values. Suppose that each node has a threshold of $\phi^* = 1/33$. In each of our simulations, we use a single node, chosen uniformly at random, as a seed at $t=0$ and consider $\tau \in \{1,2,3,4,5, 6\}$ and $p_{\mathrm{Hip}}\in [0,1.0]$ (in increments of $0.01$). 

\begin{figure*}[ht!]
\centering
\includegraphics[width=.43\linewidth]{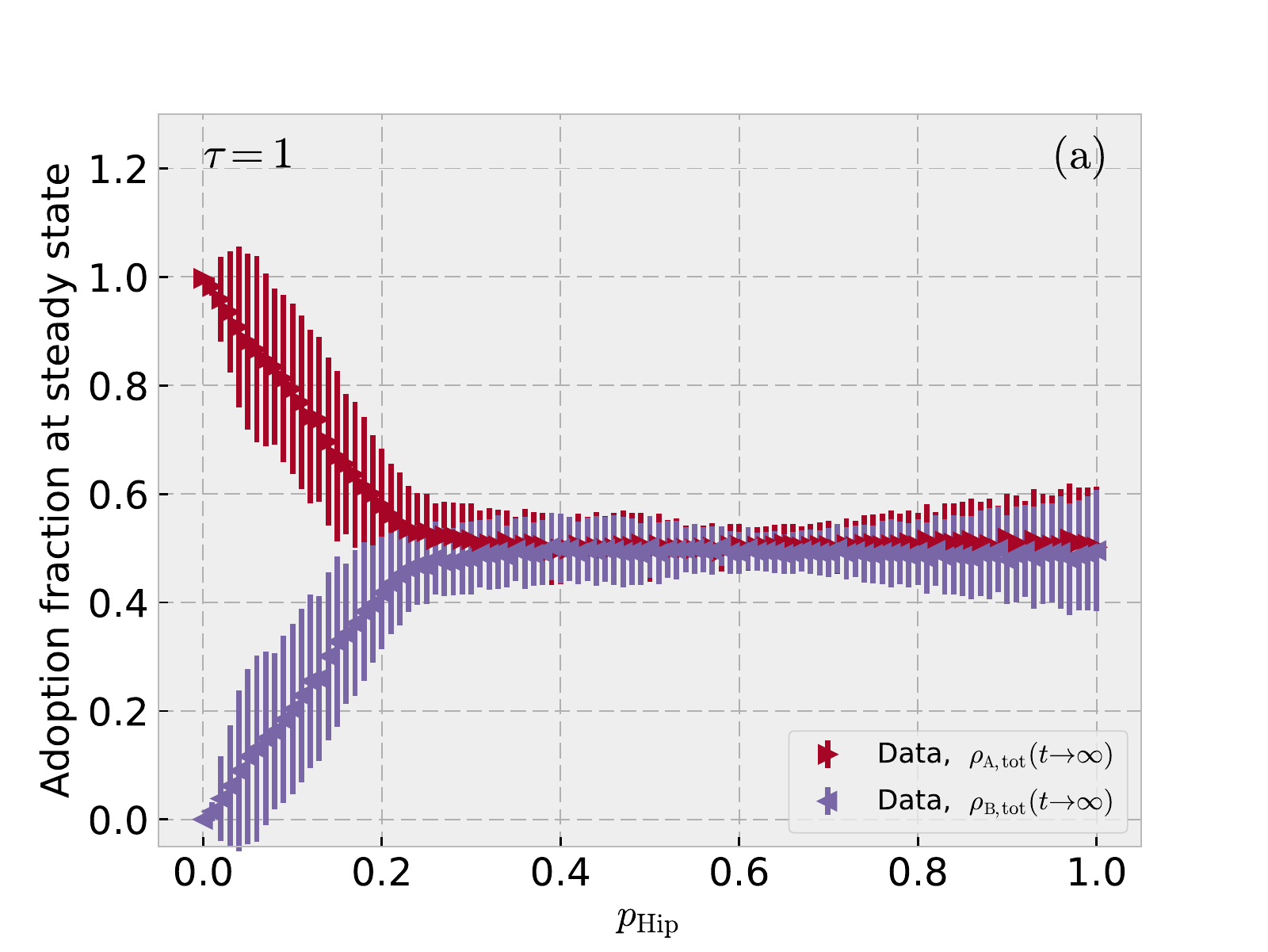}
%\hfill
\includegraphics[width=.43\linewidth]{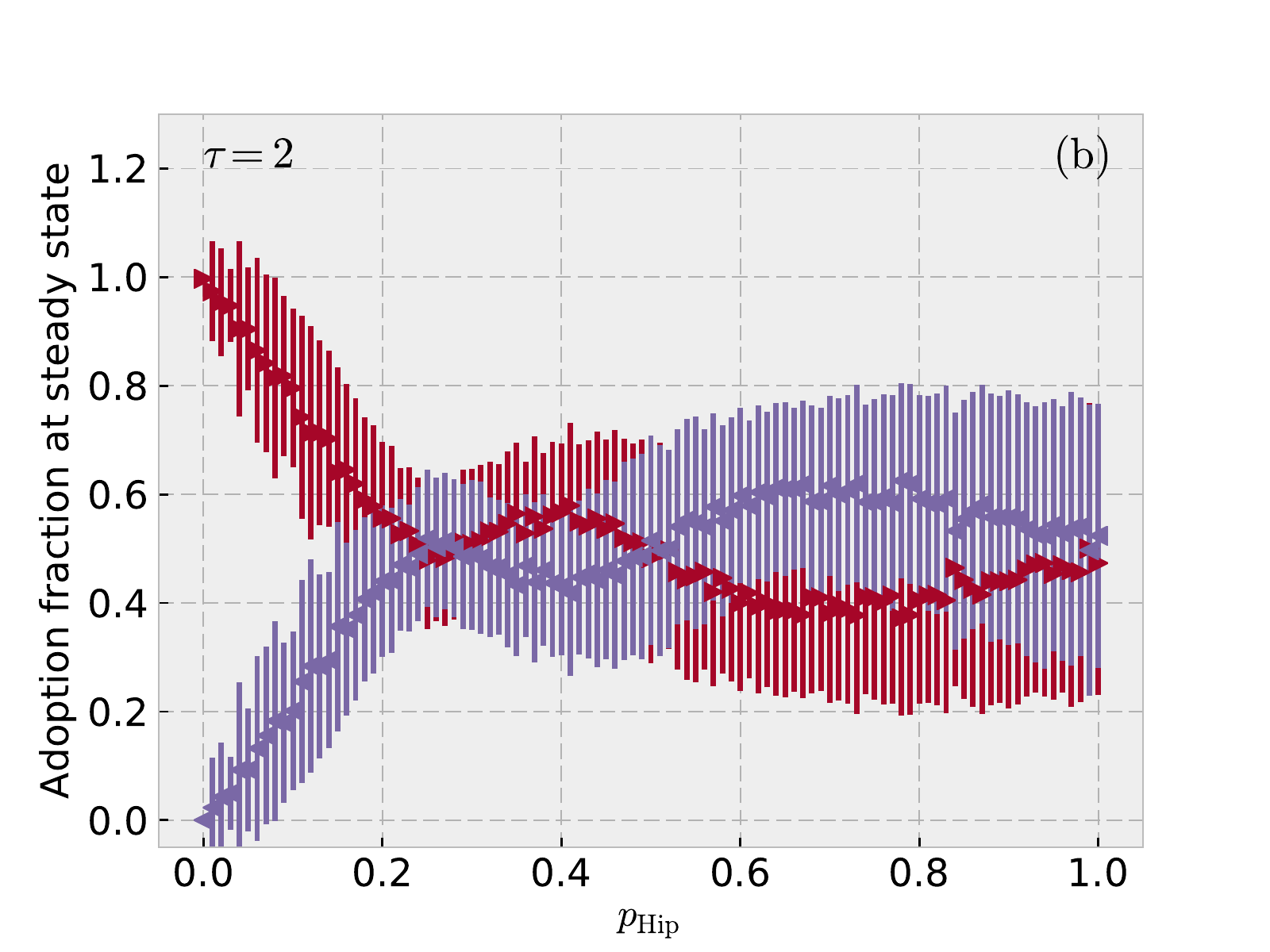}
\\
\includegraphics[width=.43\linewidth]{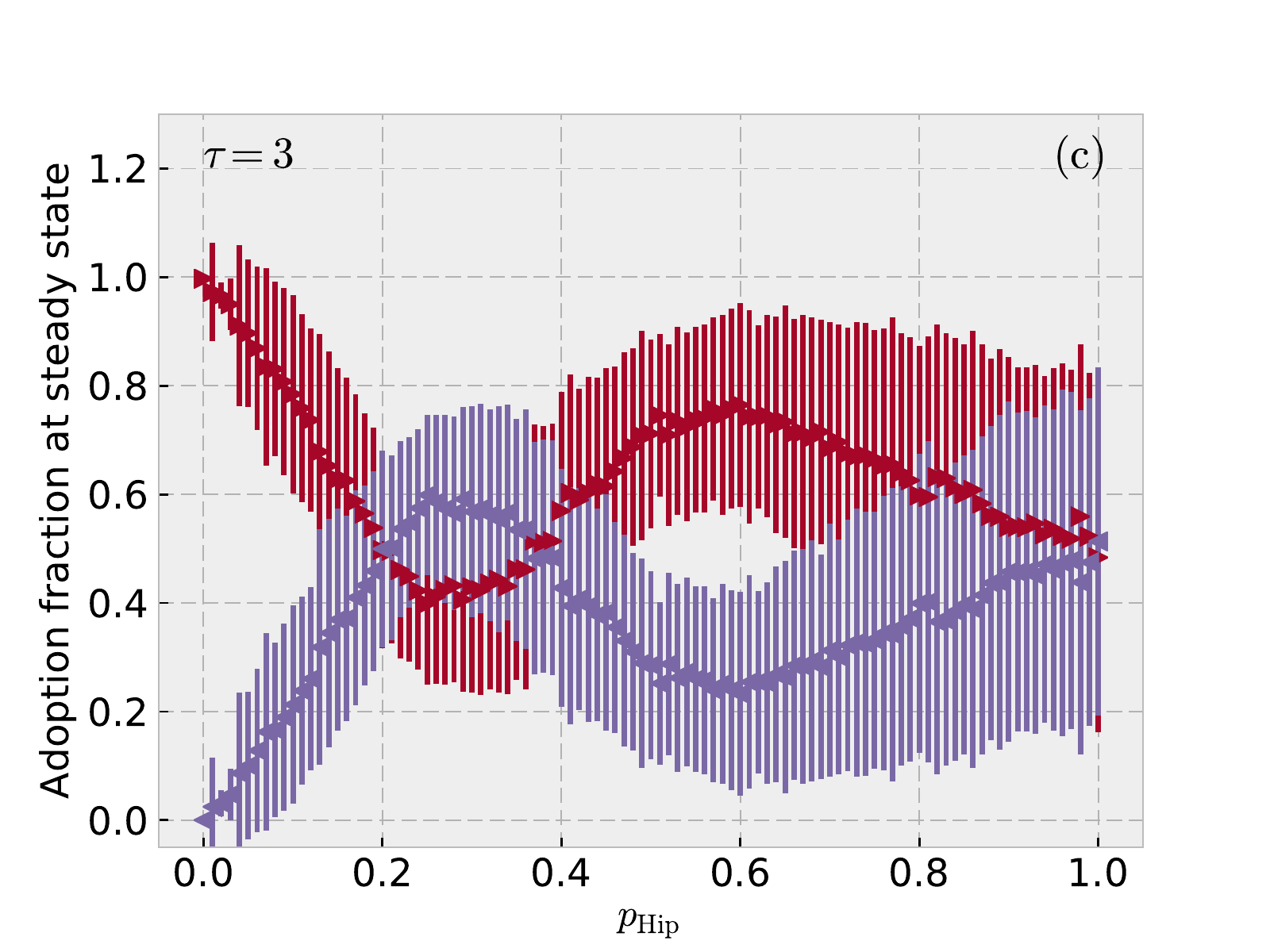}
%\hfill
\includegraphics[width=.43\linewidth]{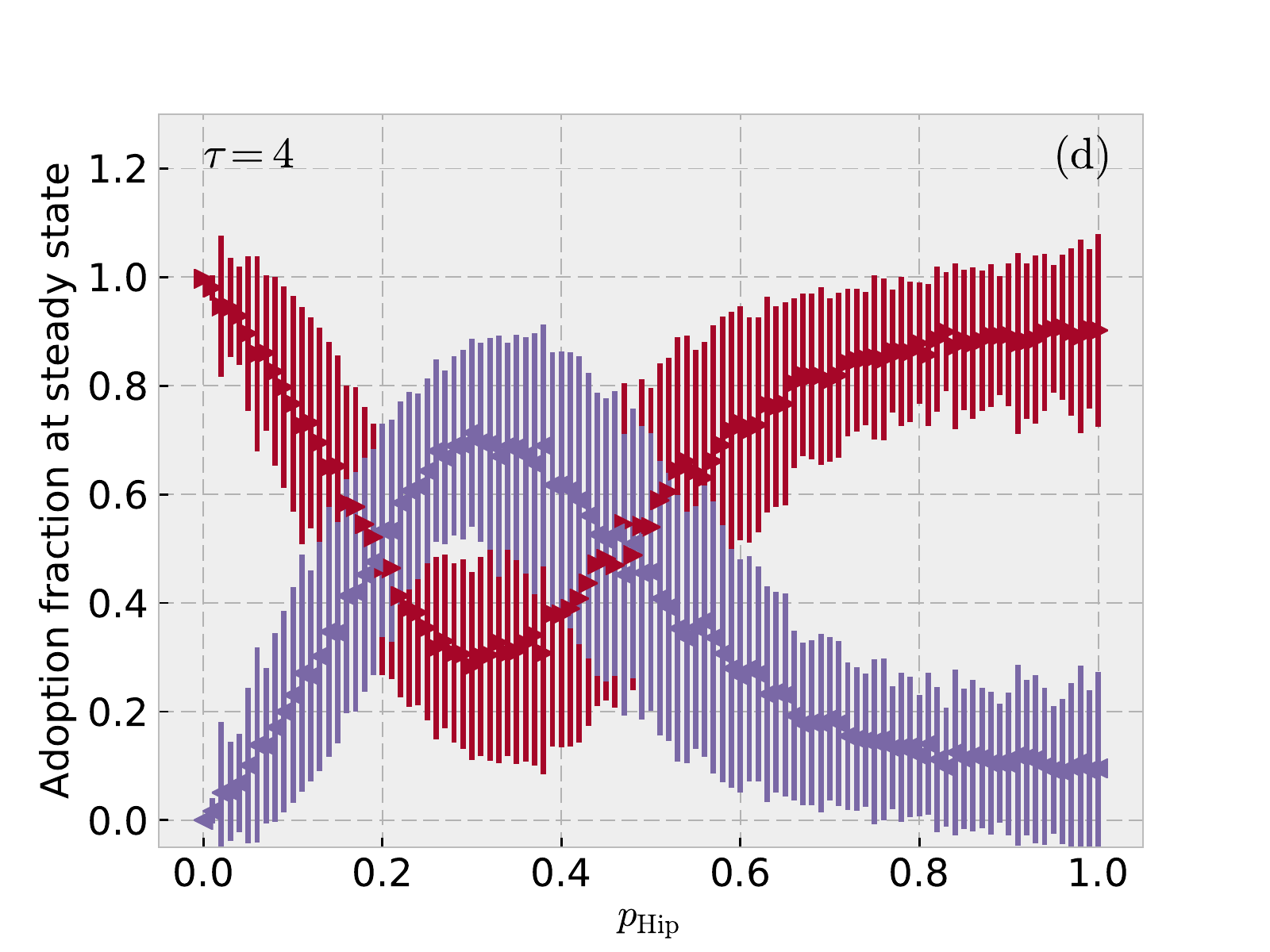}
\\
\includegraphics[width=.43\linewidth]{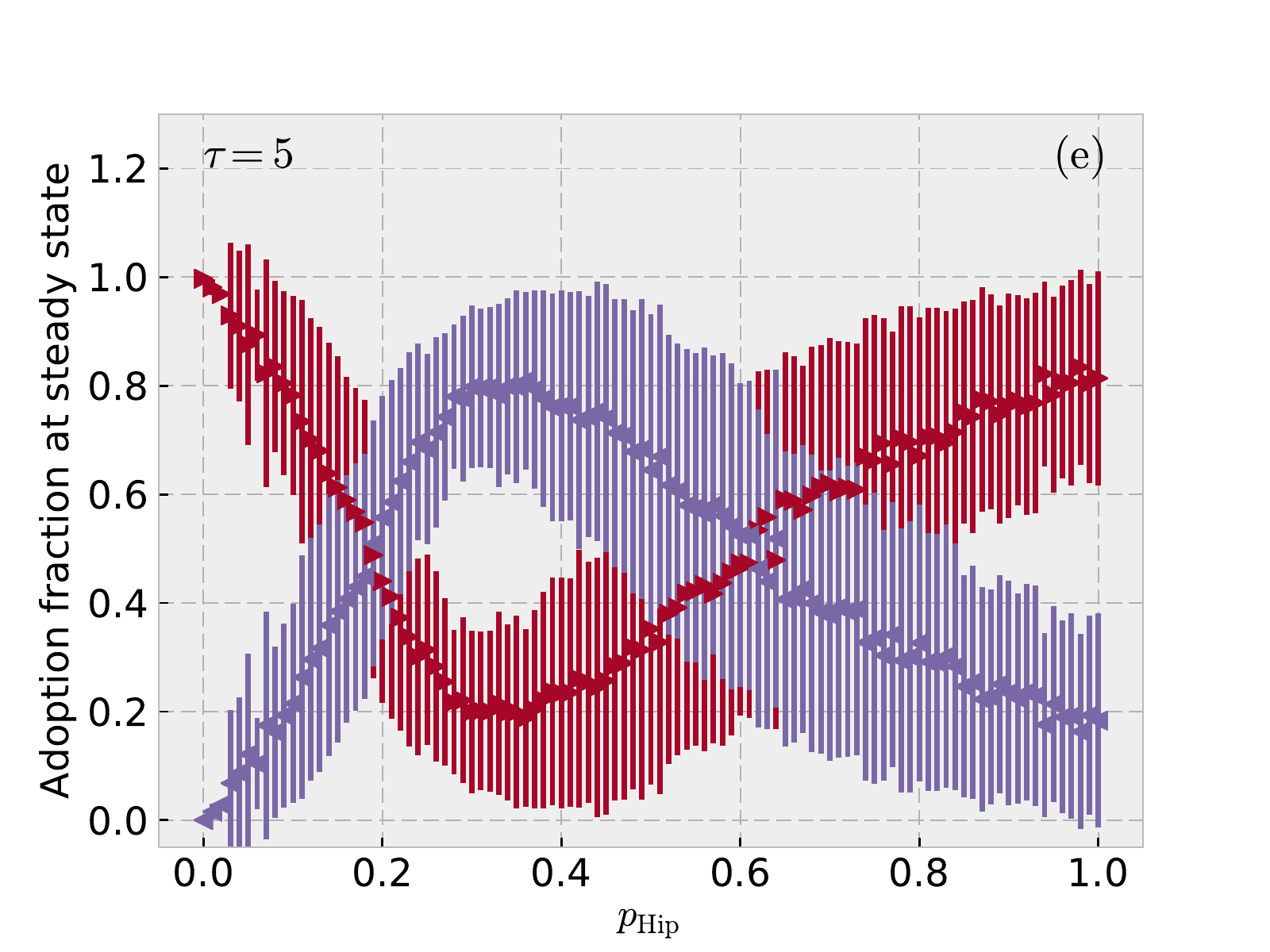}
%\hfill
\includegraphics[width=.43\linewidth]{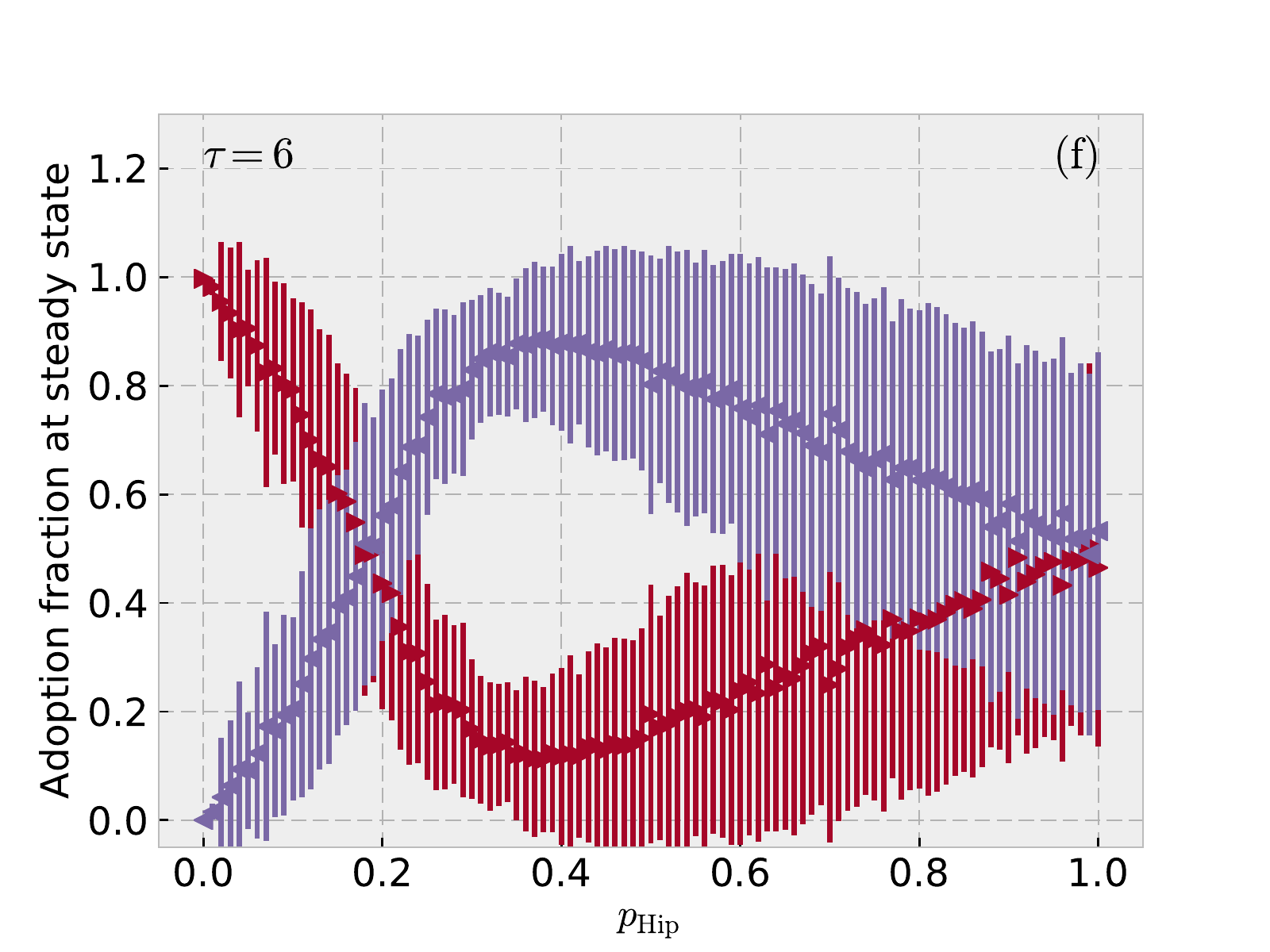}
\caption{Distribution of products at steady state for the {\sc Northwestern25} network from the {\sc Facebook100} data set. The different panels give results of simulations of our hipster threshold model with different delay times $\tau$ for the hipster nodes. For each value of $\tau$, we consider hipster probabilities $p_{\mathrm{Hip}} \in [0,1]$ in increments of $0.01$. For each $(\tau, p_{\mathrm{Hip}})$ parameter pair, we simulate the hipster threshold model on  the {\sc Northwestern25} network with $200$ choices for the seed node, chosen uniformly at random, which adopts product $A$ at $t = 0$. We use a different set of $200$ nodes for different parameter values. Each node has a threshold of $\phi = 1/33$. We plot the mean fractions of nodes that have adopted products $A$ and $B$ at steady state in the $200$ realizations and the corresponding standard deviations of the means. For all values of $\tau$, the fraction of nodes that have adopted product $B$ increases rapidly with $p_{\mathrm{Hip}}$ for small $p_{\mathrm{Hip}}$, reaching $0.5$ at $p_{\mathrm{Hip}} \approx 0.2$ for $\tau \ge 3$ [see panels (c)--(f)] and for larger values of $p_{\mathrm{Hip}}$ for $\tau \leq 2$ [see panels (a) and (b)]. For $\tau = 1$, which we show in panel (a), hipsters have information about the product distribution in the network without any delay, and the fraction of nodes that adopt products $A$ and $B$ are very similar for $p_{\mathrm{Hip}}\gtrapprox 0.3$. For larger values of $\tau$ [see panels (b)--(f)], the fraction of nodes that adopt each product varies nonmonotonically for $p_{\mathrm{Hip}}\gtrapprox 0.2$. For $\tau \geq 3$ [see panels (c)--(f)], the fraction of nodes that adopt product $B$ is largest for a small interval of $p_{\mathrm{Hip}}$ around $p_{\mathrm{Hip}} \approx 0.3$. This is the single peak in the adoption of product $B$ in the mean over these simulations. For $\tau = 2$ [see panel (b)], product $B$ is the more-popular product for large values of $p_{\mathrm{Hip}}$. For $\tau \ge 2$ [see panels (b)--(f)], product $B$ is the more-popular product for a $p_{\rm Hip}$ interval starting at $p_{\mathrm{Hip}}\approx 0.20$. The length of this interval increases with $\tau$, and both the hipster probability that produces the peak fraction in this interval and (especially) the value of the peak fraction increase with $\tau$. For $\tau = 6$ [see panel (f)], the maximum fraction of nodes that adopt product $B$ is about $0.90$.
  }
\label{fig:FB_all}
\end{figure*}

In Fig.~\ref{fig:FB_all}, we show the mean fraction of nodes that have adopted products $A$ and $B$ at steady state. For each choice of parameters, we choose a set of $200$ initial conditions, and we calculate means over these simulations. The most striking difference between these plots compared to those for our model on synthetic networks in previous sections is that now it takes more hipsters for equal fractions of nodes to adopt the two products at steady state. In the {\sc Northwestern25} network, the fractions that have adopted the two products become equal when roughly one fifth of the nodes are hipsters. We also observe that the height of the first peak of $\rho_{B,{ \mathrm{tot}}}(t\to\infty)$ increases with $\tau$, as was also the case in the synthetic networks that we examined, and the standard deviations are once again large for most parameter pairs. The mean fraction of nodes that adopt some product in the discarded realizations is $0.0001$, which again is much less than the threshold of $0.10$.

\begin{table*}[tbh!]
\begin{tabular}{l | c | c | c}
\hline
\hline
Network & Number of discarded realizations & Mean & Standard deviation of the mean \\
\hline
$5$-regular configuration model & $36$ & $0.0001$ & $0.0000$ \\ 
%\hline
$3$-regular configuration model & $1843$ & $0.0002$ & $0.0001$ \\ 
%\hline
\ER \,\,($G(N,p)$) & $52214$ & $0.0072$ & $0.0010$ \\ 
{\sc Northwestern25} & $35171$ & $0.0001$ & $0.0001$ \\ 
\hline
\hline
\end{tabular}
\caption{Summary statistics of the discarded realizations of our hipster threshold model on each network family (or individual network, for {\sc Northwestern25}). The second column gives the total number of discarded realizations. In it, we sum the instances from all parameter values, because the values of $\tau$ and $p_{\rm Hip}$ do not influence which fraction of nodes activate in a given realization. We show the mean fraction of nodes that are activated at steady state for discarded realizations and the standard deviation of this mean. For all networks, the mean fraction of active nodes is much smaller than the threshold fraction of $0.10$, below which we discard realizations. For each choice of parameter values and network, we keep $200$ realizations for our samples.
}
\label{tb:discarded}
\end{table*}

%%%%%
\section{Major impact of few individuals: Approximation on $k$-regular trees}\label{sec:regular_trees}

In Section \ref{sec:Realization}, we observed that even just a few hipster nodes can cause product $B$ to become the more-popular product at steady state, but we have not yet explored how this phenomenon can occur. In this section, we argue why even just a few anti-establishment nodes can have a major impact on steady-state adoptions in our model. From studying this mechanism, we expect some similar qualitative phenomena to occur in many other models, including ones with stochastic update rules.

\begin{figure}
\centering
\includegraphics[width=\linewidth]{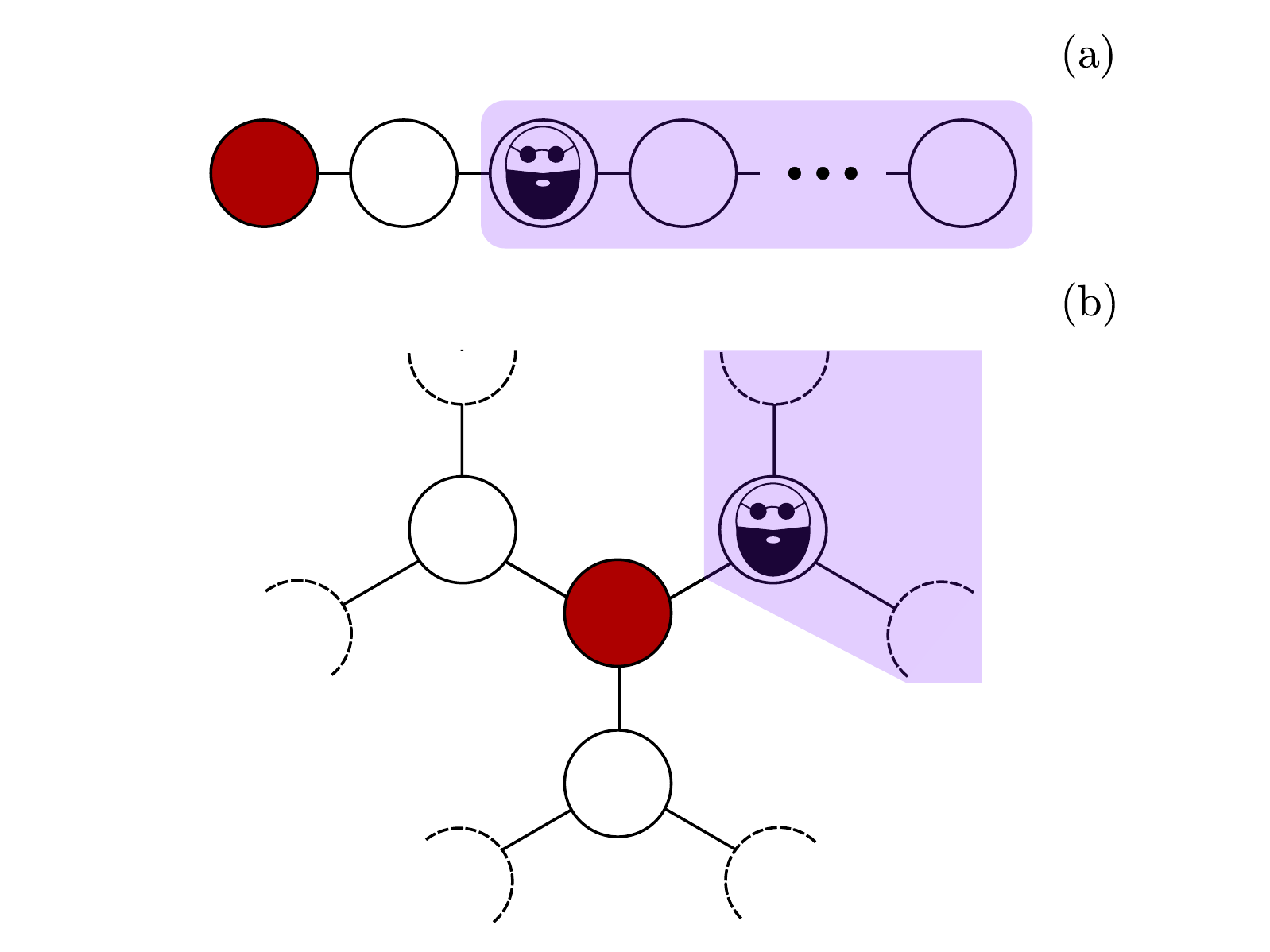}
\caption{(a) A line graph in which the leftmost node is the only seed. (It adopts product $A$.) If all nodes are vulnerable and exactly one node is a hipster, all nodes to the hipster's right eventually adopt product $B$. (b) A $3$-regular tree in which the central node is the only seed. (It adopts product $A$.) If the tree has a single hipster, all nodes that are descendants of the hipster eventually adopt product $B$.
}
\label{fig:regular_trees_illustration}
\end{figure}

To understand why even a few hipster nodes can dramatically increase the number of product-$B$ adopters at steady state, we first consider a line of $N$ nodes, which we number from one end to the other with the labels $0,1,2,\ldots,N-1$. Each node is adjacent to its immediate neighbors, with $2$ neighbors each, except for nodes $0$ and $N-1$ (which have degree $1$). For the sake of the argument, we assume that all nodes are vulnerable and that there is no delay in information (so that $\tau = 1$). We also suppose that node $0$ is the only seed, so it has adopted product $A$ at time $t=0$ (see Fig.~\ref{fig:regular_trees_illustration}a). If there are no hipsters in the line, all nodes in this scenario eventually adopt product $A$. If, by contrast, a single node $i$ is a hipster, then all nodes $j\ge i$ eventually adopt product $B$. Thus, if each node has the same independent probability of being a hipster, the expected fraction of product $B$ adopters at steady state approaches $1/2$ as $N\to \infty$. In this case, the presence of a single hipster node increases the expected fraction of nodes who adopt product $B$ at steady state from $0$ nodes to half of the nodes. The main idea is that early adopters can influence later adopters in a way that depends on the adoption paths that are available \cite{oh2018}. Moreover, although the expected fraction of product-$B$ adopters is $1/2$, a realization will be equally likely to result in any number of product-$B$ adopters, because every node is equally likely to be the hipster. This may be a reason why we observe large standard deviations in different realizations of our model on the various types of network. For more complicated network topologies, although it is no longer true in general that different steady-state fractions of product-$B$ adopters are equally probable, the steady-state adoption fraction in a given realization depends heavily on where hipsters are located in a network.

With this simple example in mind, we now turn to a more difficult example: A $k$-regular tree of vulnerable nodes in which the central node (which we label as node $0$) is the only seed (see Fig.~\ref{fig:regular_trees_illustration}b). As usual, the seed has adopted product $A$. As in the above example on a line graph, if a certain node is a hipster, it will force the nodes that follow it on an adoption path to adopt product $B$, rather than product $A$. We can divide the tree into hierarchical ``levels'': the central node is $0$, and it is adjacent to $k$ nodes in level $1$. Each node in level $1$ is adjacent to $k-1$ nodes in level $2$, each node in level $2$ is adjacent to $k-1$ nodes in level $3$, and so on. Hence, level $l\ge 1$ includes $n_l = k(k-1)^{l-1}$ nodes, and all nodes except those in the last level (which have degree $1$) have degree $k$. Such a $k$-regular tree with $L$ levels has $N=1+\sum_{l=1}^L k(k-1)^{l-1}$ nodes. In the limit of infinitely many levels, a $k$-regular tree is a Bethe lattice.

Suppose that there is a single hipster in the network. By construction, we can view any hipster as the root in a rooted tree. We can then make the following approximation. If all nodes have an equal, independent probability of being a hipster, the probability for there to be a hipster in level $l$ is equal to the fraction of nodes ($n_l/N$) that are in that level. If a hipster is present in level $l$, all nodes in an adoption path of that hipster (i.e., all of its descendants) eventually adopt product $B$. Because level $l$ has $n_l$ nodes, a single hipster in level $n_l$ causes $1/n_l$ of the nodes in later levels ($l'\ge l+1$) to adopt product $B$. Therefore, one can approximate the expected steady-state fraction of product-$B$ adopters as
\begin{align}
	\bar{\rho}_{B}(n_{\rm Hip}=1) &\approx \sum_{\lambda = 1}^L \frac{n_\lambda}{N} \frac{1}{n_\lambda}\left({\frac{1}{N}}\sum_{l=\lambda}^L k(k-1)^{l-1}\right) \notag \\
	&={\frac{1}{N^2}}\sum_{\lambda = 1}^L \sum_{l=\lambda}^L k(k-1)^{l-1}\,.
\label{eq:linear_approximation}
\end{align}
For a spreading process on a network, one can construct a dissemination tree, which describes how a contagion spreads through the network \cite{oh2018}. For a $k$-regular tree with only vulnerable nodes, the dissemination tree is the same $k$-regular tree, except that all edges are directed from the center towards the periphery. The above analysis indicates that the fraction of nodes that a single hipster can cause to adopt product $B$ is related to the properties of a dissemination tree. For dissemination trees with a progressively larger number of mean descendants per node, we expect a progressively larger fraction of nodes in an associated network to adopt product $B$ when a single hipster is present in the network. Equation~\eqref{eq:linear_approximation} illustrates that, for a given network, increasing the number of hierarchical levels in a dissemination tree tends to result in a number of product-$B$ adopters from a single hipster.

To obtain a naive estimate of the fraction of product-$B$ adopters as a function of $n_{\rm Hip}$ when $n_{\rm Hip}\ll N$, we multiply Eq.~\eqref{eq:linear_approximation} by $n_{\rm Hip}/N$, thereby assuming that adding a second hipster to the network leads to as many product-$B$ adopters as the original hipster \footnote{The original hipster was in the network before it was popular.}. However, the second hipster may be a descendant of the existing hipster, such that it does not cause any additional product-$B$ adoptions. To account for this, we develop a recursive formula that takes this possibility into account.

Imagine adding hipsters to a network one at a time (allowing the possibility of choosing the same node multiple times when attempting to add hipsters). We seek to approximate the expected fraction of product-$B$ adopters at steady state in a network with $n_{\rm Hip}$ hipsters as a function of the expected fraction of product-$B$ adopters at steady state in a network with $n_{\rm Hip}-1$ hipsters. Let $P_{\rm desc}$ denote the probability that the additional hipster is a descendant of another hipster in the network.  Adding a hipster has two possible outcomes: (1) the hipster is a descendant of another hipster, such that it does not yield additional product-$B$ adopters; or (2) the hipster is not a descendant of another hipster, so on average it yields another $\bar{\rho}_B(n_{\rm Hip}=1)$ fraction of product-$B$ adopters at steady state. We summarize this reasoning in the formula
\begin{equation}\label{hipp}
\begin{split}
	\bar{\rho}_{B}( n_{\rm Hip}) &\approx \bar{\rho}_B(n_{\rm Hip}-1)P_{\rm desc} + \large[ \bar{\rho}_{B}(n_{\rm Hip}-1) \\
&\quad+ \bar{\rho}_B(n_{\rm Hip}=1) \large] (1-P_{\rm desc})\,.
\end{split}
\end{equation}
In a $k$-regular tree, all descendants of a hipster are product-$B$ adopters at steady state, so the probability that the $n$\textsuperscript{th} hipster descends from one of the previous $n-1$ hipsters equals the expected fraction of nodes that are product-$B$ adopters at steady state in a network with $n_{\rm Hip }=n-1$ hipsters. We thus insert $P_{\rm desc} = \bar{\rho}_B(n_{\rm Hip}-1)$ into equation \eqref{hipp} to obtain
\begin{align}
	\bar{\rho}_{B}( n_{\rm Hip}) \approx \bar{\rho}_{B}( n_{\rm Hip}-1) + \left[1- \bar{\rho}_{B}( n_{\rm Hip}-1)\right]\bar{\rho}_{B}(n_{\rm Hip}=1) \,.
\label{eq:recursive_approximation}
\end{align} 

\begin{figure}
\centering
\includegraphics[width=\linewidth]{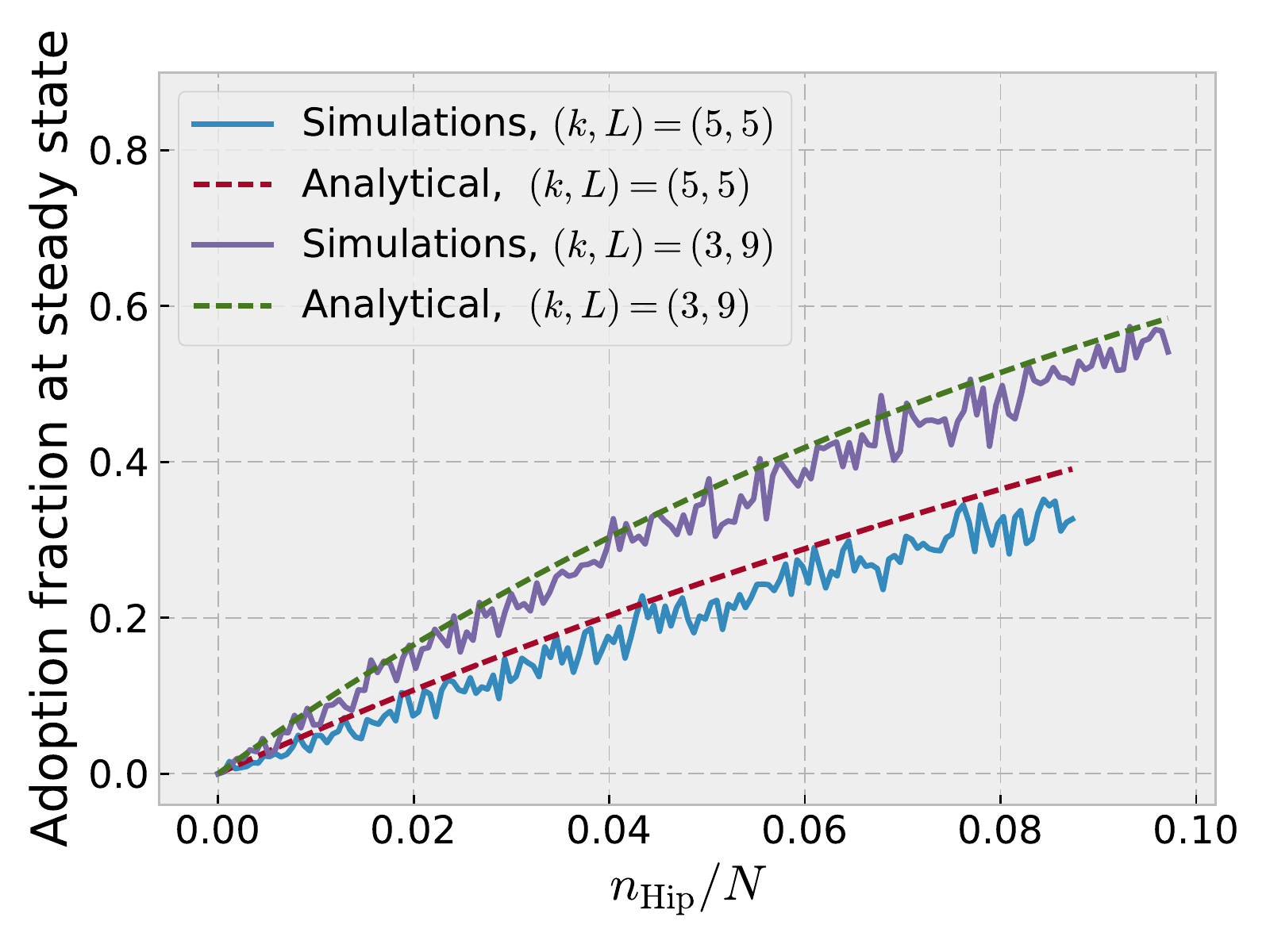}
\caption{Fraction of nodes in $k$-regular trees with $L$ levels that have adopted product $B$ at steady state. We show results for $3$-regular trees with $9$ levels (and hence with $N=1534$ nodes in total) and for $5$-regular trees with $5$ levels (and hence with $N=1706$ nodes in total).  We plot the recursive approximation from Eq.~\eqref{eq:recursive_approximation} and show our simulation results, averaged over $100$ realizations, for the steady-state fraction of nodes that have adopted product $B$ as a function of the fraction $n_{\rm Hip}/N$ (with $n_{\rm Hip}\in \{0,1,\ldots,150\}$) of hipsters in the network. As expected, our approximation is good for $n_{\rm Hip} \ll N$.
}
\label{fig:regular_trees}
\end{figure}

In Fig.~\ref{fig:regular_trees}, we compare the analytical expression in equation \eqref{eq:recursive_approximation} to computations using $3$-regular and $5$-regular trees. Our analytical approximation is a good match for our simulations when $n_{\rm Hip}/N$ is small. For larger $n_{\rm Hip}/N$, \eqref{eq:recursive_approximation} overestimates the steady-state fraction of nodes that have adopted product $B$. Hipsters need not always adopt product $B$; with more hipsters, it becomes increasingly likely that product $B$ will not always be the less-popular product.

Our analysis has several interesting consequences. For example, it yields some understanding of how the delay $\tau$ affects the steady-state adoption fractions each product. To illuminate the impact of $\tau$, it is helpful to consider the following situation. Suppose that, because of hipsters, product $B$ becomes more popular than product $A$ at some point during a simulation of our model. A delay of $\tau \ge 2$ postpones the time at which any hipsters start adopting product $A$, so we expect hipsters who adopt product $A$ to have fewer descendants than if $\tau = 1$. This provides some argument as to why the height of the peak of the fraction of product-$B$ adopters as a function of $p_{\rm Hip}$ increases with the delay, and it sheds some light on the effects of the delay. If there is no delay (i.e., $\tau = 1$) and there are many hipsters, then hipsters tend to balance the popularities of the two products, leading to roughly equal fractions of the two products at steady state (as we saw in our simulations on all networks in Section \ref{sec:Realization}).

Our analysis also improves our understanding of how various changes to our hipster model can affect steady-state results.  
For example, suppose that we use a stochastic updating rule instead of a deterministic one. Although the above analysis does not rely on the deterministic nature of our updating rule, it does indicate that adoption order is important, and anything that changes the adoption order (such as using a stochastic update rule or updating node states asynchronously instead of synchronously) may change the outcome of simulations \cite{fennell2016}. However, from our analysis, we do expect some features of our results to be robust even with different update rules and update orders. For example, for either stochastic update rules or asynchronous updating, we expect an increase in the number of steady-state product-$B$ adopters with increasing $p_{\rm Hip}$ for small values of $p_{\rm Hip}$, followed by a decrease (or stall) in the number of steady-state product-$B$ adopters when enough hipsters are present in the network (as some of them will now adopt product $A$). However, the rate at which the steady-state fraction of product-$B$ adopters increases with $p_{\rm Hip}$ for small $p_{\rm Hip}$ is likely to be influenced by stochastic update rules and asynchronous updating. For instance, suppose that we use the same rules for product selection but that we employ an asynchronous updating process in which, during each time step, we select a node uniformly at random to update, we repeat this selection process some number of times during the same time step, and we then advance time by one step. We then continue with this process in our simulations until no further spreading occurs. In this case, every node in a network can potentially adopt a product even in the first time step, and the process tends to spread at a different rate --- it can be either faster or slower --- than in synchronous updating. Because the hipsters are distributed uniformly at random and the rules governing product choice are the same as in our original model, changing the number of adopters during each time step can directly affect hipsters, as their product choice is time-dependent. (Other nodes are affected indirectly, as they can experience a different product selection in their neighborhoods.) Consequently, a faster initial spreading would increase early product-$B$ adoption for a delay $\tau \ge 2$, as all hipsters are guaranteed to choose product $B$ for time steps $t\le \tau$. To test this, we simulate the spreading of products on $5$-regular configuration-model networks with $\tau=2$, with a single product-$A$ adopter as a seed. Averaging our results over $100$ realizations (which we determine as in Fig.~\ref{fig:z_all}b), we find that the steady-state product-$B$ adoption fraction increases faster as we increase $p_{\rm Hip}$ for small values of $p_{\rm Hip}$ than what was the case for synchronous updating (see Fig.~\ref{fig:z_all}b). 
As we expected, we also find that the product-$B$ steady-state adoption fraction decreases as we increase $p_{\rm Hip}$ for larger values of $p_{\rm Hip}$. More generally, different update mechanisms and update orders can yield different dissemination trees, which describe how a contagion spreads through a network \cite{oh2018}. This can, in turn, impact steady-state product popularities.

Another aspect that tends to alter a dissemination tree is changes in the threshold distribution of nodes in a network. For example, with a threshold distribution in which all nodes are vulnerable, a spreading process can reach a steady state very quickly, and there are then few hierarchical levels in the associated dissemination tree. By contrast, a threshold distribution for which a network starts with fewer vulnerable nodes may take longer to reach a steady state, and one thus expects more levels in an associated dissemination tree. From our analysis, we see that this in turn can increase adoptions of product $B$ for small values of $p_{\rm Hip}$. Performing simulations on $5$-regular configuration model networks with $\tau = 2$ (as in Fig.~\ref{fig:z_all}b) with $p_0=1.00$, $p_0 = 0.90$, $p_0 = 0.80$, and $150$ seed nodes supports this intuition. When we examine small values of $p_{\rm Hip}$, the steady-state fraction of product-$B$ adopters increases slightly more slowly for larger values of $p_0$ as we increase $p_{\rm Hip}$. 

For some network families, we expect that networks with different numbers of nodes will have different fractions of product-$B$ adopters at steady state. To illustrate this point, we consider line graphs and $k$-regular trees. For a line graph with a single hipster and a seed node that adopts product $A$ at one end, the expected steady-state fraction of product-$B$ adopters is roughly $1/2$ for a line with any number of nodes. However, adding another level to a $k$-regular tree with a single seed node that adopts product $A$ affects the expected steady-state fraction of product-$B$ adopters. For example, a $3$-regular tree with $3$ levels (and hence with $10$ nodes in total) has an expected steady-state fraction $\bar{\rho}_B = 15/100$ of product-$B$ adopters, whereas a $3$-regular tree with $4$ levels (and hence with $22$ nodes in total) has $\bar{\rho}_B=1/10$. This difference occurs because adding another level to the $3$-regular tree increases the fraction of nodes that are leaves. Therefore, the randomly distributed hipsters have fewer descendants on average in the dissemination tree, decreasing the expected fraction of product-$B$ adopters at steady state. Using configuration-model and ER networks, we have simulated some examples with $10^3$ and $4\times 10^4$ nodes, and we observe the same qualitative results as what we present previously. More generally, however, our analysis demonstrates that the number of nodes in a network can affect steady-state product distributions. Even taking seed-size scaling into consideration (see \cite{gleeson07}), dissemination trees can still change, potentially affecting qualitative steady-state results.
 
Changing the way that nodes choose which product to adopt can also drastically influence simulation outcomes. For example, consider a modification of our model in which a hipster that becomes active at time step $t$ adopts the product that is less popular among its neighbors at time $t-\tau$. Further, suppose that two competing products are spreading in a $k$-regular tree, in which the central node is the only seed. As usual, the seed has adopted product $A$. When we constructed our approximation \eqref{eq:recursive_approximation} for the steady-state distribution of products in the limit of few hipsters, we assumed that every hipster adopts product $B$. In the modified hipster model, this approximation may be very bad. Hipsters who descend from other hipsters may adopt product $A$. We thus expect the product-$B$ steady-state adoption fraction to increase more slowly with $p_{\rm Hip}$ for small $p_{\rm Hip}$ if a hipster adopts the product that is less popular among its neighbors, rather than the less-popular product among all active nodes in the network. Performing simulations of the modified model on $5$-regular configuration model networks with $p_0=0.80$, $\tau = 1$, and a single seed node (as in Fig.~\ref{fig:z_all}a), we find that a smaller (or equal, for $p_{\rm Hip}=0$) steady-state fraction of nodes adopts product $B$ for $p_{\rm Hip}\le 0.15$, compared to our observations for our focal hipster model.

%%%%

\section{Conclusions}\label{sec:Conclusions}

It is important to study what makes information, opinions, diseases, memes, products, misinformation, alternative facts, and other things that originate in a small subpopulation spread to a large fraction of nodes in a network. Such scenarios can arise in the adoption of products and the spreading of memes; and they can also occur in anti-establishment behavior, which can significantly impact the geopolitical landscape.

We developed a threshold model to examine the impact of anti-conformists (so-called ``hipsters'') on the spreading of two competing products (one of which, labeled ``$B$'', is not adopted by any node at the beginning of our simulations). We examined our hipster threshold model on various types of networks, and we considered different fractions of the hipster nodes and different amounts of time delay in the global information that the hipsters possess. In the absence of a time delay, we found that hipsters tend to balance the adoption of the two competing products. For all other delay values and all types of networks that we examine, we observed that the fraction of nodes that adopt product $B$ (i.e., the product that would not be adopted in the absence of hipsters) grows rapidly with the fraction of hipsters. Surprisingly, for all of our networks, we needed only a small fraction of hipsters for product $B$ to become comparably popular, or even more popular, than product $A$ at steady state. In our simulations on a variety of synthetic networks, we found that it is often sufficient for fewer than 10\% of the nodes to be hipsters for product $B$ to become at least as widespread as product $A$ (the only product that spreads at time $t=0$). For the {\sc Northwestern25} Facebook network, roughly $20\%$ of the nodes need be hipsters for product $B$ to be as widespread as product $A$ at steady state. 

Using a line graph and $k$-regular trees, we illustrated why the fraction of nodes who adopt product $B$ increases rapidly for small values of $p_{\rm Hip}$. On these networks, we obtained good agreement between simulations and an approximation of the fraction of nodes who adopt product $B$ at steady state in the limit of few hipsters. From our analytical approximation in the few-hipster regime, we observed that product-$B$ adoption fraction increases with the distance between the seed node and other nodes. This gives some insight into why there is a much slower increase in product-$B$ adopters for the spreading process on the Facebook network than in synthetic networks, as the former has a smaller mean geodesic (i.e., shortest) path length than our synthetic networks. It also suggests that different realizations with identical parameter values may result in very different steady-state product adoption fractions, given that we use random processes to choose hipsters and seed nodes. One consequence of such sensitivity to initial conditions is large standard deviations in the mean steady-state adoption fraction of each product, which is what we observe in most cases. The same mechanistic insight suggests that a larger delay $\tau$ results in more hipsters adopting product $B$ early in a simulation, and each of these early adopters influences the product choice of later adopters. We believe that postponing the time at which hipsters choose product $A$ instead of product $B$ is the main reason that a progressively larger delay $\tau$ results in a progressively larger peak of the expected product-$B$ adoption fraction as a function $p_{\rm Hip}$. Finally, the mechanistic insight from our approximation in the few-hipster limit of also helps illustrate that the properties --- such as threshold distributions, the number of nodes in a network, and update rules --- of an update rule or network that affect dissemination trees (which describe how a contagion spreads through a network) can affect observations at steady state, although some qualitative observations should be robust under such variations. 

 Our hipster threshold model exhibits a variety of fascinating dynamics on different types of networks. For example, when there is a delay in global information (i.e., $\tau \geq 2$) and the hipster probability $p_{\rm Hip}$ is large, we observed nontrivial dynamics in the number of intervals of hipster probabilities for which a given product is more popular at steady state. The quality of the match between our pair approximation and numerical simulations also depends on both network structure and hipster probability. For example, our approximation was effective for small values of $p_{\rm Hip}$ and it correctly produced a fast increase in product-$B$ adopters with increasing values of small $p_{\rm Hip}$,
it did reasonably well for large values of $p_{\rm Hip}$ for $5$-regular configuration-model networks (except for abrupt jumps that are not present in the simulations), it had mixed results for $3$-regular configuration-model networks (although it yielded the correct result for the more-popular product at steady state for $p_{\rm Hip}\approx 1$ in all but one instance), and it was not good for \ER{} networks (where it was incorrect about which product is more popular at steady state for $p_{\rm Hip} \approx 1$ in roughly half of the cases).

If there is a delay (i.e., $\tau \geq 2$) in the global adoption information that is available to hipsters, we found that the steady-state fraction of nodes that adopt a product varies nonmonotonically with the fraction of hipsters. For some delay values, this steady-state fraction peaks for multiple, disparate values of the fraction of hipsters; for other delay values, however, there is only a single peak. This behavior also depends on the network type on which spreading occurs. If there is no delay in the global adoption information that is available to hipsters (i.e., $\tau = 1$), 
 we found that the final fraction of nodes that adopt product $B$ first increases rapidly with $p_{\rm Hip}$ and then stabilizes such that approximately half of the nodes adopt each product.

In summary, in our model, even when only one of two products is adopted when spreading begins, very small fractions of anti-establishment nodes can lead to a competing product being adopted by a majority of a population. Our simple model and numerical experiments may help shed light on the road to success for anti-establishment choices in elections and competition between products, as such success (and qualitative differences in final outcomes between competing products, political candidates, and so on) can arise rather generically from a small number of anti-establishment individuals and ordinary processes of social influence on normal individuals. In our model, the hipsters always choose to adopt the product that is less popular at time step $t-\tau$. If all hipsters regard product $A$ as the established choice at all time steps --- regardless of the actual distribution of adopted products --- the steady-state adoption fractions of product $B$ become even larger, and the anti-establishment choice (which is product $B$, in our example) becomes even more successful than what we observed in our simulations. This more extreme situation may be relevant in elections in which the conception of who is part of the establishment may not change during weeks of campaigning and polls predicting which candidate will win an election and take office.

In future work, it would be interesting to study our hipster model in more detail, including investigating whether the fraction of hipsters is connected to any criticality, and to extend the model in various ways. Extensions of our model may be helpful for studying the impact of anti-establishment hubs, such as alt-right broadcasting services or alt-right Twitter accounts with many followers. Understanding what makes a large voter population vulnerable to the views of a few ``hipster'' nodes may help guard populations from manipulation and fake information during elections and other scenarios.

%%%%%

\section{Acknowledgements}

Part of this work was carried out at the Mathematical Institute at University of Oxford. We thank James Gleeson, Kameron Decker Harris, Shankar Iyer, and Mikko Kivel\"a for helpful discussions; and MAP thanks Ben Williamson for early inspiration to study hipsters. We also thank Serge Galam for bringing his work to our attention. JSJ also thanks the Mathematical Institute at University of Oxford for hospitality, Mogens H. Jensen (Niels Bohr Institute, University of Copenhagen) for making this project possible, and funding through University of Copenhagen, UCPH 2016 Excellence Programme for Interdisciplinary Research, and the Danish Council for Independent Research. Vive la r\'{e}sistance! 

%%%%%%%

%\bibliography{bibliography7}

%%%%


%%%%%%

\end{document}